\documentclass[aps,prd,epsf]{revtex4}
\usepackage[latin1]{inputenc}
\usepackage{float}
\usepackage{graphicx}
\newcommand{\be}{\begin{equation}}
\newcommand{\ee}{\end{equation}}
\newcommand{\bea}{\begin{eqnarray}}
\newcommand{\eea}{\end{eqnarray}}

\newcommand{\tl}{\tilde{\ell}}

\def\bo{\kern1pt\vbox{\hrule height
1.2pt\hbox{\vrule width 1.2pt\hskip 3pt
\vbox{\vskip 6pt}\hskip 3pt\vrule width 0.6pt}\hrule
height 0.6pt}\kern1pt}

\def\delvd{\dot{\delta}_m^{(v)}}

\def\ty_3{\tilde{y}_3}

\newcommand{\rd}{{\rm d}}

\begin{document}

%____________________________________________________________________________
\title{{\large Density perturbations in $f(R)$ gravity theories \\
in metric and Palatini formalisms}}

%______________________________________
%______________________________________

\author{Shinji Tsujikawa\footnote{shinji@nat.gunma-ct.ac.jp}}
\affiliation{Department of Physics, Gunma National College of
Technology, Gunma 371-8530, Japan}

\author{Kotub Uddin\footnote{k.uddin@qmul.ac.uk}}
\affiliation{School of Mathematical Sciences,\\
Queen Mary, University of London,
London E1 4NS, UK}

\author{Reza Tavakol\footnote{r.tavakol@qmul.ac.uk}}
\affiliation{School of Mathematical Sciences,\\
Queen Mary, University of London,
London E1 4NS, UK}

%______________________________________
\begin{abstract}

We make a detailed study of matter density perturbations
in both metric and Palatini formalisms. Considering
general theories whose Lagrangian density is a general
function, $f(R)$, of the Ricci scalar $R$, we
derive the equation of matter density perturbations
in each case, in a number of gauges,
including comoving, longitudinal and uniform density gauges.
We show that for viable $f(R)$ models that satisfy cosmological
and local gravity constraints (LGC), matter perturbation equations derived
under a sub-horizon approximation are valid even for
super-Hubble scales provided the oscillating mode
(scalaron) does not dominate over the matter-induced mode.
Such approximate equations are especially reliable
in the Palatini formalism because of the absence of scalarons.

Using these equations we make a comparative study of the behaviour
of matter density perturbations as well as gravitational potentials
for a number of classes of $f(R)$ theories.
In the metric formalism the quantity $m=Rf_{,RR}/f_{,R}$
that characterises the deviation from the $\Lambda$CDM
model is constrained to be very small during a matter era
in order to ensure compatibility with LGC, but the models
in which $m$ grows to the order of $10^{-1}$ around
the present epoch can be allowed.
These models also suffer from an additional fine tuning
due to the presence of scalaron oscillating modes
which are absent in the Palatini case.

In Palatini formalism LGC and background cosmological constraints
provide only weak bounds on $|m|$ by constraining it to be
smaller than $\sim 0.1$. This is in contrast to
matter density perturbations which, on galactic scales,
place far more stringent constraints on the
present deviation parameter $m$ of the order of
$|m| \lesssim 10^{-5}$--$10^{-4}$.
This is due to the peculiar evolution of matter perturbations
in the Palatini case which exhibits a rapid growth or a damped
oscillation depending on the sign of $m$.

\end{abstract}

%-----------------------------------%
\maketitle

%%%%%%%%%%%%%%%%%%%%%%%%%%%%%%%%%%%%%%%%%%%%%%%%%%%%%%%%%%%%%%%%%%%%%%%%
\section{Introduction}
%%%%%%%%%%%%%%%%%%%%%%%%%%%%%%%%%%%%%%%%%%%%%%%%%%%%%%%%%%%%%%%%%%%%%%%%

Recent high--precision observations by the Wilkinson Microwave
Anisotropy Probe (WMAP) \cite{Spe06} together with
high redshift supernovae surveys \cite{Supern},
observations of large scale structure
\cite{Tegmark-etal04} and baryon acoustic
oscillations \cite{Eis05} have provided strong evidence that
the Universe is at present undergoing a phase of accelerated
expansion \cite{review,review2}.
Such an accelerating phase poses a serious
problem for cosmology since it is difficult
to explain within the usual
general relativistic framework.

Phenomenologically, the simplest way to generate such an
accelerating phase is through the addition of a
cosmological constant to the Einstein's field equations.
At a more fundamental level, however, the problem is how
to account for such a constant within a candidate theory of
quantum gravity. This has motivated a large number of alternative
models (see Ref.~\cite{review2} for a recent review).
These models can mostly be divided into two broad groups:
those involving an exotic matter source and
those modifying the gravitational sector of the theory.
An important subset of the latter are the
so called generalised gravity theories,
based on non-linear lagrangians of the form
$f(R)$, where $f$ is a general differentiable
function of the Ricci scalar $R$.
Such modifications to the (linear) Einstein-Hilbert action
typically arise in effective actions derived from string/M-theory
\cite{Buchbinder-et-al-1992,Gasperini-Venezinano-1992,Nojiri03,Vassilevich-2003}.

There are two approaches that may be taken in
studying generalised $f(R)$ theories of gravity,
depending upon the choice of the independent fields
with respect to which the action is varied.
In the first (`metric' approach) only variations
with respect to the metric are considered,
whereas in the second (`Palatini' approach)
the action is varied with respect to both the metric
and connection. Both approaches result
in identical field equations for the case of linear
Einstein-Hilbert action. In the more general cases,
with nonlinear Lagrangians $f(R)$,
however, the metric approach
results in fourth-order field equations, whereas
the Palatini variation generates a second-order system.
In this paper we consider both approaches.

Recently a number of $f(R)$ models have
been proposed in order to explain the late-time
acceleration of the Universe.
Given the large number of such models that have been
(or can potentially be) considered, an urgent task at
present is to devise stringent tests in order to reduce
the viable range of candidates.
In addition to constraints obtained by demanding the stability
of the models, we require the models to be consistent
with constraints from background cosmological
dynamics \cite{AGPT,AT}, as well as from
local gravity constraints (LGC) \cite{Olmo,Navarro,CSE,lgcpapers}.

Generally, models put forward to explain the
late-time acceleration involve infra-red corrections to
Hilbert action with negative powers of the Ricci scalar $R$.
It is interesting to note that the viability of
such models can depend upon which formalism is used
in order to derive them.
As an example consider the model $f(R)=R-\mu^{2(n+1)}/R^n$
with $n>0$ \cite{Capo03,Carroll03}
(see also Ref.~\cite{Capoluca}).
In the metric approach this model is problematic because of
the absence of the matter era \cite{APT,AGPT} as well as
the instability of perturbations associated with negative
$f_{,RR}$ \cite{Dolgov,Song}, where comma denotes
differentiation with respect to $R$.
Moreover it does not satisfy the LGC \cite{Olmo}.

On the other hand, in the Palatini approach the background cosmological dynamics
successfully generates a standard matter era followed by
a late-time acceleration \cite{FTS}.
One can also realise a correct Newtonian limit in the regime where
$R$ is much larger than $\mu^2$ \cite{Sotiri}.
For the study of cosmological and local gravitational constraints
for a range of $f(R)$ models,
see Refs.~\cite{metricpapers} in the metric case and
Refs.~\cite{palatinipapers} in the Palatini case.
See also Refs.~\cite{Hu07,Starobinsky07,Appleby,ShinjiNew}
for $f(R)$ theories that satisfy both
cosmological and local gravity constraints
in the metric approach.

Despite the importance of above constraints
in limiting the range of viable models,
it is well known that the study of the homogeneous
dynamics is not sufficient on its own to determine the
nature of the underlying theory uniquely,
given that $f(R)$ theories can always be
expressed in terms of a conventional relativistic cosmology
sourced by a perfect fluid with an
effective equation of state.
An important way to break this degeneracy is by considering
the evolution of density perturbations about the
Friedmann-Lemaitre-Robertson-Walker background.
The study of perturbations provides an important tool
in order to break this degeneracy as well as allowing
more stringent constraints to be placed on the
parameters of the models.
In fact a number of authors recently studied the evolution of
density perturbations for some $f(R)$ models to put constraints
on model parameters in metric \cite{Song,linear,Li,SPH} and
Palatini \cite{Amar,Koivisto06,BLi} formalisms.

Here, with this aim in mind, we study the evolution of density
perturbations and resulting observational consequences
in both metric and Palatini formalisms.
We first write the equations without specifying any gauges and
then write them in a number of gauges
(including comoving, longitudinal and
uniform density gauges) from which we derive
the corresponding approximate
perturbation equations under sub-horizon type
approximations. In the metric approach where the oscillating
mode (referred to as the scalaron, see below) \cite{star} is present,
this approximation can be invalid if the scalaron
is overproduced in the early Universe.
However, as long as the scalaron is sub-dominant relative to
a matter induced mode, we shall show that approximate
perturbation equations can be valid even for super-Hubble modes
in the models that satisfy LGC.
The approximation is especially reliable in the Palatini case
because of the absence of scalarons.
The simplicity of the equations derived facilitate the estimation
of the growth rate of perturbations both analytically
and numerically.

Using these equations we make a comparative study of the behaviour
of matter density perturbations in both formalisms,
for a number of classes of theories satisfying
LGC as well as background constraints.
These include viable theories recently proposed
in the literature \cite{Hu07,Starobinsky07,ShinjiNew}.
An important quantity to characterise the growth rate
of matter perturbations is $s \equiv \delta_m'/\delta_m$,
where $\delta_m$ is the density contrast on
orthogonal comoving hypersurfaces and
a prime represents a derivative with respect to
the number of e-foldings. In the standard general
relativistic (GR) case $s=1$.

In the metric formalism, the growth rate in the non-standard
regime where $\xi \equiv k^2/(a^2R)\,m \gg 1$
($a$ is the scale factor and $k$ is the comoving wave number)
is given by $s=1.186$. While this is within the current
observational bound $s \lesssim 1.5$, the difference
between spectral indices of the matter power spectrum and
the CMB spectrum can provide stronger constraints.
In the Palatini case, when $m>0$, the growth rate
$s$ grows exponentially once
the Universe enters the regime $\xi \gg 1$.
Thus the quantity $m$ can be severely constrained
using the information from the bounds on $s$.
In this paper we shall obtain constraints on the present
value of the deviation parameter $m$ as well as
model parameters for a number of $f(R)$
theories that include most
essential features required by viable models.

The plan of the paper is as follows.
In Section \ref{s2} we give a brief
review of constraints for $f(R)$ theories
in both metric and Palatini formalisms
provided by background cosmological evolution
and local gravity constraints.
In Section \ref{s3} we give a brief account of
metric perturbations and matter density perturbations
which are applicable to both metric and Palatini formalisms.
In Sections \ref{s4} and \ref{s5} we derive
the evolution equations for matter density perturbations
in metric and Palatini formalisms respectively
and in each case discuss the
constraints they provide by considering a number of
theories that are viable according
to background cosmological and local gravity constraints.
Finally we conclude in Section \ref{s6}.

In what follows we shall use units such that
$8\pi G=1$, but we restore bare
gravitational constant $G$ when it is needed.

%%%%%%%%%%%%%%%%%%%%%%%%%%%%%%%%%%%%%%%%%%%%%%%%%%%%%%%%%%%%%%%%%%%%%%%%
\section{Local gravity constraints and viable background cosmological evolution}
\label{s2}
%%%%%%%%%%%%%%%%%%%%%%%%%%%%%%%%%%%%%%%%%%%%%%%%%%%%%%%%%%%%%%%%%%%%%%%%

The classes of $f(R)$ general gravity theories we consider
have actions of the form:
\begin{eqnarray}
\label{action}
S=\frac12 \int {\rm d}^4x \sqrt{-g}\,f(R)+
S_m (g_{\mu \nu}, \psi_m) \,,
\end{eqnarray}
where $f$ is a general differentiable
function of the Ricci scalar $R$ and
$S_m$ corresponds to a matter action,
which depends upon the metric $g_{\mu \nu}$ and
matter fields $\psi_m$.
The Ricci scalar $R$ is defined by $R=g^{\mu \nu }R_{\mu \nu}$,
where the Ricci tensor $R_{\mu \nu}$ is
\begin{eqnarray}
\label{Rmunu}
R_{\mu \nu}=\partial_\lambda \Gamma^{\lambda}_{\mu \nu}
-\partial_{\mu}\Gamma^{\lambda}_{\lambda \nu}
+\Gamma^{\lambda}_{\mu \nu} \Gamma^{\rho}_{\rho \lambda}
-\Gamma^{\lambda}_{\nu \rho} \Gamma^{\rho}_{\mu \lambda}\,.
\end{eqnarray}
In the case of the metric formalism,
the connections $\Gamma^{\alpha}_{\beta \gamma}$
are the usual metric connections defined in terms
of the metric tensor $g_{\mu \nu}$. In this case
the field equations are obtained by varying
the action (\ref{action}) with
respect to $g_{\mu \nu}$ to give
\begin{eqnarray}
\label{metba1}
& &F(R)R_{\mu \nu}(g) -\frac12 f(R)g_{\mu \nu}-
\nabla_{\mu} \nabla_{\nu} F(R)+g_{\mu \nu}
\bo F(R)=T_{\mu \nu}\,,\\
\label{metba2}
& &3\bo F(R)+F(R)R-2f(R)=T\,,
\end{eqnarray}
where $F=\partial f/\partial R$.
Note that Eq.~(\ref{metba2}) corresponds to a trace of
Eq.~(\ref{metba1}).
The third and fourth terms on the left hand side of Eq.~(\ref{metba1})
come from the variation of $R_{\mu \nu}$ with respect to
$g_{\mu \nu}$.

In the case of the Palatini formalism
$\Gamma^{\alpha}_{\beta \gamma}$ and $g_{\mu \nu}$
are treated as independent variables.
Varying the action (\ref{action}) with respect to
$g_{\mu \nu}$ gives
\begin{eqnarray}
\label{Pala1}
& &F(R)R_{\mu \nu}(\Gamma) -\frac12 f(R)g_{\mu \nu}
=T_{\mu \nu}\,,
\end{eqnarray}
where $R_{\mu \nu}(\Gamma)$ is the Ricci tensor corresponding
to the connections $\Gamma^{\alpha}_{\beta \gamma}$,
which is in general different from the Ricci tensor
corresponding to the metric connections $R_{\mu \nu}(g)$.
Taking the trace of this equation, we obtain
\begin{eqnarray}
\label{trace}
F(R)R-2f(R)=T\,,
\end{eqnarray}
where $R(T)=g^{\mu \nu}R_{\mu \nu}(\Gamma)$ is directly
related to $T$ and is different from the Ricci scalar
$R(g)=g^{\mu \nu}R_{\mu \nu}(g)$ in the metric case.
Taking the variation with
respect to the connection, and using
Eq.~(\ref{Pala1}), we find
\begin{eqnarray}
\label{Pala3}
R_{\mu \nu}(g) -\frac12 g_{\mu \nu} R(g)
=\frac{T_{\mu \nu}}{F}
-\frac{FR(T)-f}{2F}g_{\mu \nu}
+\frac{1}{F}(\nabla_{\mu} \nabla_{\nu}F
-g_{\mu \nu} \bo F)
-\frac{3}{2F^2} \left[ \partial_{\mu}F \partial_{\nu}F
-\frac12 g_{\mu \nu} (\nabla F)^2 \right]\,.
\end{eqnarray}
%
%The left hand side is associated with the quantities that
%are evaluated in terms of $g_{\mu \nu}$,
%whereas all the terms on the right
%hand side are expressible in terms of $R(T)$.

It is useful to express the generalised
theories based on action (\ref{action})
in terms of generalised scalar-tensor theories
of Brans-Dicke type with a scalar field $\phi$ and the
corresponding potential $V(\phi)$ thus \cite{Chiba03}:
\begin{eqnarray}
\label{action2}
S=\frac12 \int {\rm d}^4x \sqrt{-g} \left[ \phi R
-\frac{\omega_{\rm BD}}{\phi} (\nabla \phi)^2
-V(\phi) \right] +S_m(g_{\mu \nu}, \psi_m)\,,
\end{eqnarray}
where $\omega_{\rm BD}$ is the Brans-Dicke
parameter.
Note that the original Brans-Dicke
theory \cite{BD} corresponds to $V(\phi)=0$.
To see the correspondence of the action (\ref{action2})
with (\ref{action}), we recall that the variation
of (\ref{action2}) with respect to $g_{\mu \nu}$
and $\phi$ leads to the following equations
\begin{eqnarray}
\label{BD1}
& &R_{\mu \nu}(g)-\frac12 g_{\mu \nu}R(g)=
\frac{1}{\phi}T_{\mu \nu}-
\frac{1}{2\phi}g_{\mu \nu}V(\phi)+
\frac{1}{\phi}(\nabla_{\mu} \nabla_{\nu}\phi
-g_{\mu \nu} \bo \phi)+
\frac{\omega_{\rm BD}}{\phi^2}
\left[ \partial_{\mu}\phi \partial_{\nu}\phi
-\frac12 g_{\mu \nu} (\nabla \phi)^2 \right]\,,\\
\label{BD2}
& & (3+2\omega_{\rm BD})\bo \phi+2V(\phi)
-\phi V_{,\phi}=T\,.
\end{eqnarray}
Now consider the following correspondence
\begin{eqnarray}
\label{phidef}
\phi=F(R)\,,\quad V(\phi)=R(\phi)F-f(R(\phi))\,,
\end{eqnarray}
where $R=R(g)$ in the metric case and $R=R(T)$
in the Palatini case.
Comparing Eqs.~(\ref{BD1})-(\ref{BD2}) with
Eqs.~(\ref{metba1})-(\ref{metba2}),
it can readily be seen that the $f(R)$ theories
in the metric formalism correspond
to the above generalised Brans-Dicke-type
theories with $\omega_{\rm BD}=0$.
Similarly the $f(R)$ theories in the
Palatini formalism correspond to the Brans-Dicke-type
theories with $\omega_{\rm BD}=-3/2$.

The Newtonian effective  gravitational constant in the
Brans-Dicke theory (\ref{action2}) can be derived under
a weak-field approximation by considering
a spherically symmetric body with
a mass $M_{\odot}$, constant density $\rho$ and
a radius $r_{\odot}$ and a vanishing
density ($\rho=0$) outside the body.
Decomposing the field $\phi$ into background and
perturbation parts ($\phi=\phi_0+\delta \phi$)
and using a linear perturbation theory in the Minkowski
background with a perturbation $h_{\mu \nu}$,
the effective gravitational constant
is given by \cite{Olmo}:
\begin{eqnarray}
\label{effectiveG-M}
G_{\rm eff}^{{\rm Newton}}=\frac{G}{\bar{\phi}}
\left( 1+ \frac{e^{-M\ell}}{3+2\omega_{\rm BD}} \right)\,,
\quad {\rm for} \quad
Mr_{\odot} \ll 1\,.
\end{eqnarray}
Here $\ell$ is a distance from the center of the body
and the scalar field mass squared is defined by
\begin{eqnarray}
\label{masssq}
M^2 = \frac{1}{3+2\omega_{\rm BD}}
\left( \bar{\phi} \frac{{\rm d}^2V}{{\rm d}
\bar{\phi}^2}-\frac{{\rm d}V}
{{\rm d}\bar{\phi}} \right)\,,
\end{eqnarray}
where $\bar{\phi}$ is a local field in Minkowski spacetime.
We should emphasise here that the expression (\ref{effectiveG-M})
is only valid subject to $Mr_{\odot} \ll 1$ \cite{Navarro,CSE}.
If the scalar field mass $M$ is large, it can happen that the condition
for the linear perturbation theory ($\delta \phi \ll \phi_0$)
becomes invalid. Moreover, this validity depends on
the distribution of scalar field mass inside and outside the body.
When the mass in the region $\ell<r_{\odot}$ is much larger than
the corresponding mass in the region $\ell>r_{\odot}$,
a thin-shell can be formed
inside the body to satisfy local gravity constraints through
a chameleon mechanism \cite{Chameleon}.
The formation of the thin-shell occurs in a non-linear region
in which the above linear result (\ref{effectiveG-M})
ceases to be valid \cite{Navarro}.

An important point to note here is that the Palatini case,
corresponding to $\omega_{\rm BD}=-3/2$, is
rather special in a number of fundamental ways. For example,
the $\phi$ field kinetic term in Eq.~(\ref{BD2}) vanishes in this case,
whereas it is non-zero in the metric case with
$\omega_{\rm BD} =0$.
As we shall see below this has the important consequence that
the oscillatory scalaron mode is absent in
$f(R)$ theories based on the Palatini formalism,
whereas it is present in all other models with
$\omega_{\rm BD} \neq -3/2$, including
$f(R)$ theories based on the metric formalism.

Also the scalar field mass $M$ blows up for finite
potential-dependent terms in the parenthesis of Eq.~(\ref{masssq})
as $\omega_{\rm BD}$ approaches $-3/2$.
For the theories with $\omega_{\rm BD} \neq -3/2$,
the scalar field mediates a fifth force with an interaction range $1/M$
because the field has a kinetic term in Eq.~(\ref{BD2}).
In the Palatini formalism the field lacks a kinetic term in Eq.~(\ref{BD2}),
which implies that the notion of the usual interaction
range determined by mass $M$ does not hold.
Thus the Palatini case should be treated separately
compared to other theories.

In the usual Brans-Dicke theory with $V(\phi)=0$ and
$\omega_{\rm BD} \neq -3/2$, the mass $M$ vanishes
because the field $\phi$ propagates freely.
Then the Yukawa-correction term $e^{-M\ell}$ in Eq.~(\ref{effectiveG-M})
becomes one, in which case the Brans-Dicke parameter $\omega_{\rm BD}$
is constrained by local gravity experiments to be larger
than $40000$ \cite{BDcon}.
This constraint does not, however, necessarily
apply to $f(R)$ gravity
theories, because the presence of the potential $V(\phi)$ can make
such theories compatible with local gravity constraints
under certain conditions (which we shall see later).

Before proceeding to consider matter density perturbations
and the constraints arising from them, we shall first
briefly review the cosmological and
local gravity constraints for $f(R)$ theories in both
formalisms. The constraints from background dynamics
have been studied in both metric \cite{APT,AGPT,AT}
and Palatini \cite{Amar,FTS} formalisms.
To discuss these constraints,
it is useful to introduce the following dimensionless
parameters \cite{AGPT}
\begin{eqnarray}
\label{mdef}
m=\frac{Rf_{,RR}}{f_{,R}}\,,\quad
r=-\frac{Rf_{,R}}{f}\,.
\end{eqnarray}
In the metric case insight into the cosmological
dynamics can be obtained by considering
trajectories in the $(r, m)$ plane.
In the Palatini case it is more convenient to use
other dimensionless variables for the background
dynamics \cite{FTS}, but the deviation parameter
$m$ is still important in order to characterise the
deviation from the $\Lambda$CDM model.

%--------------------------------
\subsection{Metric formalism}
%---------------------------------

We first review the cosmological viability in the metric
case and then proceed to consider the LGC.
In the $(r,m)$ plane the matter point corresponds to
$P_M: (r,m) \approx (-1,0)$.
In order to have a saddle matter era followed by
a late-time acceleration, we require the following
conditions \cite{AGPT}
\begin{eqnarray}
\label{con0met}
m>0\,,\quad -1<\frac{{\rm d}m}{{\rm d}r}<0\,,\quad
{\rm at} \quad (r,m) \approx (-1,0)\,.
\end{eqnarray}
De-Sitter fixed points $P_A$ lie on the line $r=-2$.
They are stable provided that
\begin{eqnarray}
\label{con2met}
r=-2\,, \quad 0<m \le 1\,.
\end{eqnarray}
As long as conditions (\ref{con0met}) and (\ref{con2met})
are satisfied, we can realize a successful cosmological sequence
(radiation, matter, and de-Sitter epochs).

There are a number of models in the recent literature
that satisfy the above cosmological
constraints. Examples are
(i) $f(R)=\alpha (R^b-\Lambda)^c$ with $c \ge 1$,
$bc \approx 1$ \cite{Li}, and
(ii) $f(R)=R-\alpha R^{\beta}$ with $\alpha>0$ and
$0 <\beta<1$ \cite{AT}.
For these models, the parameters $m$ and $r$
satisfy the relation $m=C(-r-1)$, where $C$ is a positive constant
in the neighbourhood of $r=-1$.
Using observational constraints on the background dynamics from SN Ia and the
sound horizon of CMB, the parameter $m$ has been
shown to be constrained to be $m<{\cal O}(0.1)$ ~\cite{AT}.

If the information from LGC is also included
the constraints on the model parameters become very strong.
The usual procedure to determine the local gravity constraints
for $f(R)$ theories is to consider their Brans-Dicke
representations (\ref{phidef})
and expand the equations of motion around
a background Minkowski metric \cite{Olmo}.
Since the metric formalism corresponds to
$\omega_{\rm BD} =0$, the scalar field mass defined in
Eq.~(\ref{masssq}) is given by
\begin{eqnarray}
\label{Massdef}
M^2=\frac13 \left( \frac{f_{,R}}{f_{,RR}}-R \right)\,,
\end{eqnarray}
where we have used the relation (\ref{phidef}).
If $M^2<0$ the Yukawa correction $e^{-M\ell}$ is replaced by an oscillating
function $\cos (|M|\ell)$, but this case is excluded
by the experimental requirement that $\gamma \approx 1$.
Hence the mass squared $M^2$ is required to be positive.

Clearly we require the mass $M$ to be heavy
for consistency with local gravity experiments.
In that case, however, the effective gravitational constant
(\ref{effectiveG-M}) obtained under the linear approximation
ceases to be valid. As was already mentioned above,
a thin-shell begins to form
through a chameleon effect in this non-linear regime.
To consider this chameleon effect in $f(R)$ gravity, it is
convenient to introduce a new conformally related
metric and a scalar field \cite{Maeda}:
\begin{eqnarray}
\tilde{g}_{\mu \nu}=\phi g_{\mu \nu}\,, \quad
\varphi=\sqrt{3/2}\,{\rm ln}\,\phi\,.
\end{eqnarray}
Then the action (\ref{action}) in the Einstein frame
takes the form:
\begin{equation}
\label{actione}
S=\int{\rm d}^{4}x\sqrt{-\tilde{g}}\left[
\frac{1}{2}\tilde{R}-\frac12 (\tilde{\nabla}\varphi)^2
-U(\varphi) \right]
+S_m(\tilde{g}_{\mu \nu}
e^{2\beta \varphi}, \psi_m)\,,
\end{equation}
where the coupling $\beta$ in $f(R)$ models
and the potential $U$ are given by
\begin{equation}
\beta=-\frac{1}{\sqrt{6}}\,,\quad
U=\frac{R(\phi)\phi-f}{2\phi^2}\,.
\end{equation}

In a spherically symmetric setting with an energy density $\rho~(=-T)$,
the field $\varphi$ satisfies the following equation \cite{Faul}
\begin{eqnarray}
\label{dreq}
\frac{{\rm d}^2 \varphi}{{\rm d} \tl^2}+
\frac{2}{\tl} \frac{{\rm d}\varphi}{{\rm d}\tl}=
\frac{{\rm d}V_{\rm eff}}{{\rm d}\varphi}\,,
\end{eqnarray}
where $\tl$ is the distance from the center
of symmetry in the Einstein frame and
\begin{eqnarray}
\label{Veff}
V_{\rm eff}(\varphi)=U(\varphi)+
e^{\beta \varphi}\rho^{*}\,.
\end{eqnarray}
Here we have introduced an energy density
$\rho^{*} \equiv e^{3\beta \varphi}\rho$, which
is conserved in the Einstein frame \cite{Chameleon}.

As an example, let us consider a spherically symmetric body
that has an energy density $\rho^{*}=\rho_A^{*}$
inside the body ($\tl<\tilde{r}_{\odot}\equiv
e^{-\beta \varphi}r_\odot$).
Let the energy density outside the body ($\tl>\tilde{r}_{\odot}$)
be given by $\rho^{*}=\rho_B^{*}$, which is much smaller
than $\rho_A^{*}$.
Then the effective potential (\ref{Veff}) has two minima at
$\varphi = \varphi_A$ and $\varphi = \varphi_B$
satisfying the relations
\begin{eqnarray}
&&U_{,\varphi} (\varphi_A)+\beta e^{\beta \varphi_A}\rho_A^{*}=0\,,\\
&&U_{,\varphi} (\varphi_B)+\beta e^{\beta \varphi_B}\rho_B^{*}=0\,.
\end{eqnarray}
The effective mass at the potential minima are given by
$m_A^2 \equiv V_{{\rm eff}}''(\varphi_A) \gg
m_B^2 \equiv V_{{\rm eff}}''(\varphi_B)$,
which comes from the condition $\rho_A^{*} \gg \rho_B^{*}$.

Imposing appropriate boundary conditions at
$\tl=0$ and $\tl=\tilde{r}_{\odot}$,
the solution to Eq.~(\ref{dreq}) can be approximated
by \cite{Chameleon,Navarro,Faul}
\begin{eqnarray}
\label{phir1}
\varphi(\tl) \simeq -\frac{\beta_{\rm eff}}{4\pi}
\frac{M_\odot e^{-m_B(\tl-\tilde{r}_\odot)}}{\tl}+\varphi_B\,,
\end{eqnarray}
where $M_\odot=4\pi r_\odot^3\rho_A/3
=4\pi \tilde{r}_\odot^3\rho_A^{*}/3$,
\begin{eqnarray}
\beta_{\rm eff} = 3\beta
\frac{\Delta \tilde{r}_\odot}
{\tilde{r}_\odot}\,,\quad
\frac{\Delta \tilde{r}_\odot}{\tilde{r}_\odot}
= \frac{\varphi_B-\varphi_A}
{6\beta \Phi_\odot}\,,
\label{phiBA}
\end{eqnarray}
and $\Phi_\odot=GM_\odot/\tilde{r}_\odot$.
A thin-shell is developed under the condition
$\Delta \tilde{r}_\odot/\tilde{r}_\odot \ll 1$.
In this case the effective coupling $|\beta_{\rm eff}|$ becomes
much smaller than $|\beta|$ so that the models can be consistent
with local gravity constraints \cite{Chameleon}.

Models that can satisfy these conditions have
recently been proposed
by (i) Hu \& Sawicki \cite{Hu07} and (ii)
Starobinsky \cite{Starobinsky07}:
\begin{eqnarray}
\label{model1}
& &{\rm (i)}~~f(R)=R-\lambda R_c \frac{(R/R_c)^{2n}}
{(R/R_c)^{2n}+1}\,, \\
\label{model2}
& &{\rm (ii)}~~f(R)=R-\lambda R_c \left[ 1-
\left( 1+\frac{R^2}{R_c^2} \right)^{-n} \right]\,,
\end{eqnarray}
where $n, \lambda$ and $R_c$ are positive constants.
In both models the cosmological constant
disappears in a flat spacetime, i.e., $f(R=0) =0$.
Note that $R_c$ is roughly of the order of the
present cosmological Ricci scalar $R_0$ for
$n={\cal O}(1)$ and $\lambda={\cal O}(1)$.
In high curvature regimes
$R \gg R_c$ these models behave as
\begin{equation}
\label{asypo}
f(R) \simeq R-\lambda R_c \left[
1-\left( \frac{R_c}{R}\right)^{2n} \right]\,,
\end{equation}
with
\begin{equation}
\label{our model}
m \simeq C(-r-1)^{2n+1}\,,
\end{equation}
where $C$ is a positive constant, and $m$ and $r$
are defined in Eq.~(\ref{mdef}).
Thus they are very close to the $\Lambda$CDM model
with suppressed values of $m$ during matter and radiation
eras ($r \simeq -1$).

In the regimes $R \gg R_c$
one can show that the term $|\varphi_B-\varphi_A|$
in Eq.~(\ref{phiBA}) is of the order of $m(R_B)$
for $n={\cal O}(1)$, where $R_B$ is the Ricci scalar
in the neighbourhood of $\varphi_B$
(which is generally much larger than $R_c$
in an environment where local tests of gravity are carried out).
Hence the thin-shell is developed under the condition
\begin{equation}
\label{mcons}
m(R_B) \ll \Phi_\odot\,.
\end{equation}
This can be regarded as a criterion
for the compatibility with local gravity constraints.
In the case of the earth, the condition (\ref{mcons})
corresponds to $m(R_B) \ll \Phi_\odot \sim 10^{-9}$.
Since $\Phi_\odot \ll 1$ in most of local gravity
experiments, the parameter $m$ is constrained
to be much smaller than the order of unity
in the region where the Ricci scalar $R_B$
is much larger than the present
cosmological one ($R_0 \sim R_c$).

Cosmologically the condition (\ref{mcons}) implies that
viable models need to be very close to the
$\Lambda$CDM model in radiation and matter dominated
epochs ($R \gg R_0$).
However, deviations from the $\Lambda$CDM model
are allowed around the present acceleration epoch ($R \sim R_0$).
Thus in viable models the parameter $m$ is negligibly small
during the radiation and matter eras, but continues to grow
by the present epoch.

For the theories of the type (\ref{asypo})
the corresponding Brans-Dicke field $\phi$,
the potential $V(\phi)$
and the mass squared $M^2$ are given by
\begin{eqnarray}
& & \phi \simeq 1-2n\lambda (R_c/R)^{2n+1}\,,\\
& & V(\phi) \simeq \lambda R_c \left[1-(2n+1) \left( \frac{1-\phi}
{2n \lambda} \right)^{\frac{2n}{2n+1}} \right]\,, \\
& & M^2 \simeq \frac{R_c}{3(2n+1)}
(2n\lambda)^{\frac{1}{2n+1}}
(1-\phi)^{-\frac{2n+2}{2n+1}}\,,
\end{eqnarray}
which in the limit $R/R_c \to \infty$ become $\phi \to 1$,
$V(\phi) \to \lambda R_c$ and $M^2 \to \infty$
respectively.
In these regimes the field is stuck around $\phi=1$
because of the presence of a $\rho$-dependent term.
When $R$ decreases to the order of $R_c$, the field begins to
evolve along the potential $V(\phi)$ with a lighter
mass $M$ which is not very much different from $R_c$.
Thus in the Brans-Dicke description,
the departure from the point $\phi=1$ amounts to
deviation from the $\Lambda$CDM model.

The models (\ref{model1}) and (\ref{model2})
are constructed to satisfy the stability conditions
\begin{eqnarray}
\label{con3met}
f_{,R}>0\,,\quad f_{,RR}>0\,,\quad
{\rm for}~~R>R_1\,(>0)\,,
\end{eqnarray}
where $R_1$ is a Ricci scalar at the late-time de-Sitter point.
The first condition is required to avoid repulsive gravity, whereas
the second ensures the absence of tachyons or ghosts.
The second condition is also required for the consistency with LGC
(as was shown above) as well as to ensure the stability of
density perturbations \cite{linear,Song} (as we shall see below).
We also note that the requirements (\ref{mcons}) and (\ref{con3met})
are perfectly consistent with the condition $0<m(R) \ll 1$
derived in Ref.~\cite{AGPT}
for the existence of the matter era.

To summarise, the conditions (\ref{mcons}) and (\ref{con3met}),
together with the existence of the de-Sitter point (\ref{con2met}),
are required for the viability of $f(R)$ models in the metric formalism.
The condition for the existence of the saddle matter era given in
Eq.~(\ref{con0met}) is automatically
satisfied under the requirements (\ref{mcons}) and (\ref{con3met}).

%---------------------------------
\subsection{Palatini formalism}
%---------------------------------

Let us next consider cosmological and local gravity constraints
for $f(R)$ theories in the Palatini formalism.

We first discuss the conditions for the cosmological
viability of $f(R)$ models at the background level.
In Ref.~\cite{FTS} it was shown that
radiation ($P_r$), matter ($P_m$) and de-Sitter ($P_A$) points
exist irrespective of the forms of $f(R)$ provided that the function
\begin{eqnarray}
\label{CR}
C(R)=-3\frac{(FR-2f)F_{,R}R}
{(FR-f)(F_{,R}R-F)}\,,
\end{eqnarray}
is well-behaved (i.e., it does not show discontinuous
or divergent behaviour).
Note that effective equations of state corresponding to points
$P_r$, $P_m$ and $P_A$ are given by
$w_{\rm eff}=1/3, 0, -1$, respectively.
The de-Sitter point $P_A$ corresponds to $FR-2f=0$, i.e.,
\begin{eqnarray}
\label{Palar}
r=-2\,,
\end{eqnarray}
and $C(R)=0$.
This de-Sitter solution exists on the same line as in the metric case.
Since the eigenvalues of the Jacobian matrix
for perturbations around the point $P_d$ are
$(\lambda_1, \lambda_2)=(-3-C(R),-4-C(R))$, the
de-Sitter point on the line $r=-2$ is always a stable attractor.
This situation is different from the metric case in which
the stability of the de-Sitter point requires
the additional condition $0<m(r=-2) \le 1$.

The stability of radiation and matter points, on the other
hand, depends upon the particular $f(R)$ models chosen.
The eigenvalues for perturbations are
given by $P_r$: $(\lambda_1, \lambda_2)=(4+C(R),1)$
and $P_m$: $(\lambda_1, \lambda_2)=(3+C(R),-1)$ \cite{FTS}.
Thus the models with $C(R)>-3$ give rise to the unstable
node for $P_r$ and the saddle point for $P_m$.
Thus models satisfying the condition $C(R)>-3$
lead to a sequence of radiation, matter and de-Sitter epochs.
As an example, let us consider the following
model \cite{Capo03,Carroll03}
\begin{eqnarray}
\label{inverse}
f(R)=R-\frac{\mu^{2(n+1)}}{R^n}\,,
\end{eqnarray}
where $\mu$ and $n$ are constants.
In this case one has $C(R)=3n$ in the regime $R^{n+1} \gg \mu^{2(n+1)}$,
which means that a successful background trajectory is realised
for $n>-1$. Note that a stable de-Sitter solution exists with
$R_1^{1+n}=(2+n)\mu^{2(n+1)}$ and $C(R)=0$.
Obviously the constraints for a successful trajectory, at least at the
background level, are not so severe compared to the metric formalism.
One does not even require the condition $m>0$ for the existence of
a viable matter era.

Let us next discuss LGC in the Palatini case.
Considering the Brans-Dicke picture of
$f(R)$ theories, the theories in the Palatini formalism
correspond to $\omega_{\rm BD}= -3/2$.
Now since the usual description of the interaction range
determined by the inverse of mass $M$ can not be
applied in this case, one needs to proceed in a different way\cite{Olmo}.
{}From Eq.~(\ref{trace}) the field $\phi=F(R)$ depends upon
the value of the trace $T$, i.e., $\phi=\phi(T)$.
We expand the field around the vacuum:
$\phi(T)=\phi_0+(\partial_T \phi_0) T+\cdots$, where
$\phi_0=\phi(T=0)$ and $T \approx -\rho[1-{\cal O}(v^2/c^2)]$.
Note that we use the non-relativistic approximation under which
the velocity $v$ of matter is much smaller than the speed of light $c$.
Carrying out a post-Newtonian expansion around the Minkowski vacuum
($g_{\mu \nu}=\eta_{\mu \nu}+h_{\mu \nu}$) in the solar system,
the solutions for the second-order perturbation equations
are given by \cite{Olmo}
\begin{eqnarray}
\label{hooP}
& & h_{00}^{(2)} \simeq \frac{2G_{\rm eff}^{\rm Newton}
M_\odot}{\ell}-
\frac{V_0}{6\phi_0}\ell^2+{\rm log}\,(\phi/\phi_0)
\,, \\
\label{hijP}
& & h_{ij}^{(2)} \simeq \left[ \frac{2\gamma
G_{\rm eff}^{\rm Newton}
M_\odot}{\ell}+\frac{V_0}{6\phi_0}\ell^2
-{\rm log}\,(\phi/\phi_0) \right] \delta_{ij}\,,
\end{eqnarray}
where $V_0=V(\phi_0)$.
Here the effective gravitational constant and the post-Newtonian
parameter are
\begin{eqnarray}
G_{\rm eff}^{\rm Newton}
=\frac{G}{\phi_0}\left(1+\frac{M_V}{M_\odot}\right)\,,
\quad \gamma=\frac{M_\odot-M_V}{M_\odot+M_V}\,,
\end{eqnarray}
where $M_\odot$ and $M_V$ are given by
\begin{eqnarray}
\label{Mod}
M_\odot=\int {\rm d}^3x\, \rho(t,{\rm x})
\frac{\phi_0}{\phi}\,,\quad
M_V= \phi_0 \int {\rm d}^3x\,
\left(\frac{V_0}{\phi_0}-\frac{V}{\phi}\right)\,,
\end{eqnarray}
and $\rho$ is the energy density of the sun.

To ensure LGC, three conditions need to be satisfied \cite{Olmo}:
\begin{itemize}
\item (i) $|M_V| \ll |M_{\odot}|$\,,
\item (ii) $|V_0\,\ell^2/\phi_0| \ll 1$\,,
\item (iii) The contribution of the  term ${\rm log}\,(\phi/\phi_0)$
is negligible.
\end{itemize}

The first condition comes from the experimental requirement
$\gamma \approx 1$.
Since it is not easy to interpret this requirement directly, we shall elucidate
this by considering a specific $f(R)$ model later.
Concerning condition (ii), setting $T=0$ in
Eq.~(\ref{trace}) and using (\ref{phidef}) to obtain $V_0=f(R_0)$,
this translates into
\begin{eqnarray}
\biggl| \frac{f(R_0)}{f_{,R_0}(R_0)} \biggr| \ell^2 \ll 1\,.
\end{eqnarray}
When the deviation from the $\Lambda$CDM model is not large,
the term $f(R_0)/f_{,R_0}(R_0)$ is of the order of the present cosmological
Ricci scalar $R_0 \sim H_0^2$. Hence on the scales of the solar system
this condition is well satisfied.

Regarding condition (iii),
the presence of the term ${\rm log}\,(\phi/\phi_0)$
in Eqs.~(\ref{hooP}) and (\ref{hijP})
leads to an additional acceleration of particles that
should be small in order to be consistent with experiments.
{}From the validity of classical Euler equations, the condition (iii)
translates to \cite{Olmo}
\begin{eqnarray}
\biggl| \frac{\rho_s \partial \phi/\partial T}{\phi} \biggr| \ll 1\,,
\end{eqnarray}
where $\rho_s$ is an energy density of the local structure.
This means that the field $\phi(T)$ should
not have a strong dependence on $T$.
Using the relations $T=2V-\phi V_{,\phi}$ and
$\phi V_{,\phi \phi}-V_{,\phi}=f_{,R}/f_{,RR}-R$,
this condition translates to
\begin{eqnarray}
\biggl| \frac{\rho_s/f_{,R_s}}{f_{,R_s}/f_{,R_sR_s}-R}
\biggr| \ll 1\,.
\end{eqnarray}
It is sometimes useful to rewrite this condition
in terms of the variable $m$ thus:
\begin{eqnarray}
\label{PalaLGC}
\biggl| \frac{1}{m(R_s)}-1 \biggr| \gg
\frac{1}{f_{,R_s}} \frac{\rho_s}{R_s}\,.
\end{eqnarray}
When $|m(R_s)| \ll 1$, this is well satisfied since
both $f_{,R_s}$ and $\rho_s/R_s$ are of the order of unity.
Note that this constraint is not so restrictive compared to
the metric formalism. This can be understood by recalling that
in the Palatini case the field is non-dynamical without an
interaction range.
In the metric formalism one needs a large scalaron
mass $M$ to satisfy the thin-shell condition, which
leads to a very small value of $m(R_s)$ satisfying Eq.~(\ref{mcons}).
We also note that in the Palatini case
the condition $f_{,RR}>0$ is not required in order to satisfy LGC.

As a concrete example, let us apply the above constraints
to the theories given by Eq.~(\ref{inverse}) with $n>-1$.
In order to give rise to a late-time acceleration, $\mu$
needs to be of the order of the present Hubble radius $H_0$.
The field $\phi$ and the potential $V(\phi)$
defined in Eq.~(\ref{phidef}) are in this case given by
\begin{eqnarray}
& &\phi=1+n\left( \frac{\mu^2}{R} \right)^{n+1}\,,\\
& &V(\phi)=(n+1) \left(\frac{\mu^2}{R}\right)^n \mu^2=
(n+1)\mu^2 \left( \frac{\phi-1}{n} \right)^{n/(n+1)}\,.
\end{eqnarray}
Now in the de-Sitter case [vacuum ($T=0$)]
the solution $R_0$ satisfies
\begin{eqnarray}
F(R_0)R_0-2f(R_0)=0\,,
\end{eqnarray}
which for the model (\ref{inverse}) gives
\begin{eqnarray}
R_0=(n+2)^{1/(n+1)}\mu^2\,,
\end{eqnarray}
and
\begin{eqnarray}
\phi_0=\frac{2(n+1)}{n+2}\,,\quad
V_0=\frac{n+1}{(n+2)^{n/(n+1)}}\mu^2\,.
\end{eqnarray}
In settings where local gravity experiments are
carried out, the parameter $\epsilon \equiv \mu^2/R_s \sim \rho_0/\rho_s$
is much smaller than unity.
For example if we take the mean density
$\rho_s=10^{-11}\,$g/cm$^3$ and use the typical values
$\mu^2 \sim H_0^2 \sim \rho_0=10^{-29}\,$g/cm$^3$
and $R_s \sim \rho_s$,
then $\epsilon$ is of the order of $10^{-18}$.

When $n>0$, then in the limit $\epsilon \to 0$, we have $\phi \to 1$
and $V(\phi) \to 0$.
Thus in the expression of $M_\odot$ given in Eq.~(\ref{Mod})
the term $V_0/\phi_0$ dominates
over the term $V/\phi$, thus giving
\begin{eqnarray}
M_V \approx \int {\rm d}^3 x\, V_0
\approx \int {\rm d}^3 x\,\mu^2\,,\quad
M_\odot \approx \int {\rm d}^3 x\, \rho_s\,.
\label{Mst}
\end{eqnarray}
Now since $\mu^2 \sim \rho_0 \ll \rho_s$,
then the condition (i) is well satisfied.

When $-1<n<0$, then as $\epsilon \to 0$
one has $\phi \to 1$ and the potential $V$ becomes
of the order $V \sim \mu^2 (\mu^2/R)^n \gg V_0 \sim \mu^2$.
This gives
\begin{eqnarray}
|M_V| \approx \int {\rm d}^3 x\,\mu^2 (\mu^2/R)^n
\approx \int {\rm d}^3 x\,\rho_0 (\rho_0/\rho_s)^n\,,
\end{eqnarray}
where $M_\odot$ is the same as that in Eq.~(\ref{Mst}).
The ratio of the integrands in the expressions
for $M_V$ and $M_\odot$ can be
estimated to be $(\rho_0/\rho_s)^{n+1} \ll 1$,
which means that the condition $|M_V| \ll M_\odot$
is again satisfied.

The parameter $m$ in this case is given by
\begin{eqnarray}
m=-\frac{(n+1)n\epsilon^{n+1}}{1+n\epsilon^{n+1}}\,.
\end{eqnarray}
Now since $\epsilon$ is much smaller than 1 we obtain $|m(R_s)| \ll 1$.
Hence theories of type (\ref{inverse}) with $n>-1$ can satisfy
local gravity constraints.

The above discussion shows that
it is easier to satisfy the local gravity constraints
in the Palatini case than in the metric case.
In the latter case
we also require the condition $f_{,RR}>0$
to ensure that the scalaron mass squared $M^2$ is positive.
Moreover the requirement of the heavy mass $M$ gives
very small values for $m(R_s)$, which
imposes the condition that viable $f(R)$ models need to
be very close to the $\Lambda$CDM model
during matter and radiation epochs.
We also note that even though the condition $|m(R_s)| \ll 1$
is also required in the Palatini case, the absolute
values of $|m(R_s)|$ do not need to be vanishingly small.
In fact even models (\ref{inverse})
with $n>0$ can satisfy the correct Newtonian limit,
while they are excluded in the metric formalism
because $f_{,RR}$ is negative in those cases.
Thus in the Palatini formalism the models of
the type $f(R)=R-g(R)$ can be consistent with local gravity tests
provided that the contribution of the term $g (R)$ is
not significant relative to the linear term.

In subsequent sections we discuss the evolution of density
perturbations for $f(R)$ theories in both metric and Palatini
formalisms. We shall carry out a detailed analysis
for a number of $f(R)$ models
that can satisfy both the cosmological and local
gravity constraints and use the evolution of density
perturbations to place constraints on the model
parameters as well as their deviation from the $\Lambda$CDM model.

%%%%%%%%%%%%%%%%%%%%%%%%%%%%%%%%%%%%%%%%%%%%%%%%%%%%%%%%%%%%%%%%%%%%%%%%
\section{Matter perturbations and gauge issues}
%%%%%%%%%%%%%%%%%%%%%%%%%%%%%%%%%%%%%%%%%%%%%%%%%%%%%%%%%%%%%%%%%%%%%%%%
\label{s3}

In this section we present the equations for matter perturbations
that are applicable to $f(R)$ theories in both metric and
Palatini formalisms.
As our background spacetime we shall consider a flat
Friedmann-Lemaitre-Robertson-Walker (FLRW).
The perturbed FLRW metric which includes
linear scalar metric perturbations $\alpha$, $b$, $\varphi$ and $E$
can be written in the form \cite{metper}
\begin{eqnarray}
\rd s^2=-(1+2\alpha) \rd t^2-2a b_{,i}
\rd t \rd x^{i}+a(t)^2 \left[(1+2\varphi)
\delta_{ij}+2E_{|ij} \right] \rd x^i \rd x^j\,,
\end{eqnarray}
where $a(t)$ is a scale factor.
We shall consider a pressure-less matter
source. Now since the flow is irrotational for
scalar perturbations, we can introduce a
velocity potential $V$ in terms of which
the components of the energy
momentum tensor of the pressure-less matter
can be decomposed as
\begin{eqnarray}
T^{0}_{0}=-(\rho_m+\delta \rho_m)\,,\quad
T^0_{i}= \rho_m(V-b)_{,i} \equiv -\rho_mv_{m,i}\,,
\end{eqnarray}
where $v_m$ is related to the velocity potential $V$
through \cite{karim}:
\begin{eqnarray}
v_m=-(V-b)\,.
\end{eqnarray}
Note that the definition of $b$
in Ref.~\cite{karim} has an opposite sign to that used here.
We recall that in both metric and Palatini formalisms
the matter energy density $\rho_m$ satisfies the standard
continuity equation
\begin{eqnarray}
\dot{\rho}_m+3H\rho_m=0\,,
\end{eqnarray}
where $H=\dot{a}/a$ is the Hubble parameter and a
dot represents a derivative with respect to $t$.
The matter perturbation can then be shown to satisfy the following
equations of motion in the Fourier space \cite{HN1,Koivisto}:
\begin{eqnarray}
\label{ma1}
& & \delta \dot{\rho}_m+3H \delta \rho_m
=\rho_m \left( \kappa -3H \alpha -\frac{k^2}{a}v_m
\right)\,, \\
\label{ma2}
& & \dot{v}_m+H v_m=\frac{1}{a}\alpha\,,
\end{eqnarray}
where $k$ is a comoving wavenumber and
\begin{eqnarray}
\kappa \equiv 3(H\alpha -\dot{\varphi})+
\frac{k^2}{a^2}\chi\,,\quad
\chi \equiv a(b+a \dot{E})\,.
\end{eqnarray}
Defining the following variables
\begin{eqnarray}
\label{cove}
v \equiv a v_m=-a(V-b)\,, \quad
\delta \equiv \frac{\delta \rho_m}{\rho_m}\,,
\end{eqnarray}
where $v$ is a covariant velocity perturbation \cite{Malikwands}.
Eqs.~(\ref{ma1}) and (\ref{ma2}) can be written as
\begin{eqnarray}
\label{ma1d}
& & \alpha=\dot{v}\,, \\
\label{ma2d}
& &  \dot{\delta}=\kappa-3H\alpha-\frac{k^2}{a^2}v\,.
\end{eqnarray}

Now choosing a comoving hypersurface, the density perturbation
can be expressed in a gauge-invariant way as \cite{karim}:
\begin{eqnarray}
\delta \tilde{\rho}_m=\delta \rho_m
+a\dot{\rho}_m (V-b)\,.
\end{eqnarray}
We shall define the density contrast on comoving
orthogonal hypersurfaces as
\begin{equation}
\label{deltam}
\delta_m=\frac{\delta \rho_m}{\rho_m}+3Hv\,.
\end{equation}
Now since the right hand side of Eq.~(\ref{deltam})
is gauge-invariant, $\delta_m$ can be evaluated in any gauge.
The evolution equation for
${\delta}_m$ is given by
\begin{eqnarray}
\label{mattere}
\ddot{\delta}_m+2H\dot{\delta}_m+\frac{k^2}{a^2}
(\alpha-\dot{\chi})=3\ddot{B}+6H\dot{B}\,,
\end{eqnarray}
where $B=Hv-\varphi$.

In the following we shall consider three different gauges
to fix the gauge degree of freedom:

\begin{itemize}
\item {\bf Comoving gauge:} in which the 3-velocity and
the scalar shift function vanish (i.e., $v=0$).
This implies that along with the 3-velocity the momentum
vanishes as well. Thus the gauge-invariant
$\delta_{m}$ in this gauge becomes
\begin{equation}
\delta_m^{(v)}=\frac{\delta \rho_m}{\rho_m}{\Bigg|}_{v=0}.
\end{equation}
\item {\bf Longitudinal gauge:} in which the shift
vector $b$ and the anisotropic potential $E$ both vanish,
resulting in $\chi=0$.
The gauge-invariant $\delta_{m}$ in this gauge becomes
\begin{equation}
\label{delchidef}
\delta_m^{(\chi)}=\frac{\delta \rho_m}{\rho_m}+3Hv{\Bigg|}_{\chi=0}\,.
\end{equation}
\item {\bf Uniform density gauge:} in which we have constant density
hypersurfaces, i.e. $\delta \rho_m=0$.
The gauge-invariant $\delta_{m}$ in this gauge becomes
\begin{equation}
\delta_m^{(\delta)}=3Hv{\Big|}_{\delta \rho_m=0}\,.
\end{equation}
\end{itemize}

In what follows we discuss the evolution of matter perturbations
in the above three gauges, while noting that
the physics does not depend upon the choice of gauges.
Also while the above discussions hold in both metric and Palatini
formalisms, the equation of matter perturbations is
different in each case as we shall see in the following sections.

%%%%%%%%%%%%%%%%%%%%%%%%%%%%%%%%%%%%%%%%%%%%%%%%%%%%%%%%%%%%%%%%%%%%%%%%%%%%%%%%%%%
\section{Density perturbations in the metric formalism}
%%%%%%%%%%%%%%%%%%%%%%%%%%%%%%%%%%%%%%%%%%%%%%%%%%%%%%%%%%%%%%%%%%%%%%%%%%%%%%%%%%%%
\label{s4}

In the metric formalism the background equations
are given by
\begin{eqnarray}
\label{beM1}
& & 3FH^2=\frac12 (FR-f)-3H\dot{F}+\rho_m\,,\\
\label{beM2}
& & -2F\dot{H}=\ddot{F}-H\dot{F}+\rho_m\,,
\end{eqnarray}
where $R=6(2H^2+\dot{H})$.
Note that we only take into account a non-relativistic matter.
In Fourier space the scalar metric perturbations satisfy the
following equations of motion, in the
so-called gauge-ready form \cite{HN1}
\begin{eqnarray}
\label{met1}
& & -\frac{k^2}{a^2}\varphi+3H(H\alpha-\dot{\varphi})
+\frac{k^2}{a^2}H \chi=\frac{1}{2F} \left[
3H \delta \dot{F}-\left(3\dot{H}+3H^2-\frac{k^2}{a^2}
\right) \delta F-3H \dot{F} \alpha -\dot{F} \kappa
-\delta \rho_m \right]\,,\\
\label{met2}
& & H\alpha-\dot{\varphi}=\frac{1}{2F}
\left[ \delta \dot{F}-H \delta F-\dot{F} \alpha+\rho_m
v \right]\,, \\
\label{met3}
& & \dot{\chi}+H \chi-\alpha-\varphi=
\frac{1}{F} ( \delta F-\dot{F}\chi)\,, \\
\label{met4}
& & \dot{\kappa}+2H\kappa+\left( 3\dot{H}-
\frac{k^2}{a^2} \right) \alpha \nonumber \\
& &=\frac{1}{2F} \left[ \left(-6H^2+\frac{k^2}{a^2}
\right) \delta F+3H\delta \dot{F} +3\delta \ddot{F}
-\dot{F}\kappa-3(2\ddot{F}+H\dot{F})\alpha
-3\dot{F}\dot{\alpha}+\delta \rho_m \right], \\
\label{met5}
& & \delta \ddot{F}+3H\delta \dot{F}+\left(\frac{k^2}
{a^2} -\frac{R}{3} \right)\delta F
=\frac13 \delta \rho_m+\dot{F} (\kappa+\dot{\alpha})
+(2\ddot{F}+3H\dot{F})\alpha-\frac13 F \delta R\,.
\end{eqnarray}
In the following we derive the perturbation equations
in the above gauges, in both exact forms as well as
using a sub-horizon approximation.

%--------------------------------------
\subsection{Comoving gauge ($v=0$)}
%--------------------------------------

We first derive the equation of matter
perturbations in the comoving gauge ($v=0$).
When $v=0$ we have $\alpha=0$ and
$\delvd=\kappa$ from Eqs.~(\ref{ma1d})
and (\ref{ma2d}). Hence from Eq.~(\ref{met4}) we find
\begin{eqnarray}
\label{delm}
\ddot{\delta}_m^{(v)}+\left(2H+\frac{\dot{F}}{2F}
\right) \dot{\delta}_m^{(v)} =
\frac{1}{2F} \left[ \left(-6H^2+\frac{k^2}{a^2}
\right)\delta F+3H\delta \dot{F}+3\delta \ddot{F}
+\delta \rho_m \right]\,,
\end{eqnarray}
whereas from Eq.~(\ref{met5}), the perturbation
$\delta F$ satisfies
\begin{eqnarray}
\label{delF}
\delta \ddot{F}+3H\delta \dot{F}+\left(\frac{k^2}
{a^2} +\frac{f_{,R}}{3f_{,RR}}
-4H^2-2\dot{H} \right)\delta F
=\frac13 \delta \rho_m+\dot{F}
\dot{\delta}_m^{(v)}\,.
\end{eqnarray}

The evolution of the matter perturbations
$\delta_m^{(v)}$ can then be obtained by solving
Eqs.~(\ref{delm}) and (\ref{delF}) numerically.
For models that satisfy local gravity
constraints the mass squared term defined in
Eq.~(\ref{Massdef}) can be well approximated by
$M^2 \simeq \frac{f_{,R}}{3f_{,RR}}$, a term which appears
on the left hand side of Eq.~(\ref{delF}).
We are mainly interested in the evolution of
perturbations on sub-horizon scales, i.e.,
\begin{eqnarray}
\label{percon1}
\frac{k^2}{a^2} \gg \{ H^2, |\dot{H}| \}\,.
\end{eqnarray}
We also recall that for the models that satisfy LGC
the mass of the scalar field squared $M^2$ is much larger than
$R \sim H^2 \sim |\dot{H}|$.
Hence either $k^2/a^2$ or $M^2$ is dominant
in the parenthesis on the left hand side
of Eq.~(\ref{delF}).
Let us first consider the case in which
the time-derivative terms in $\delta F$
are neglected, i.e.,
\begin{eqnarray}
\label{percon2}
\left\{ \frac{k^2}{a^2}|\delta F|,
M^2 |\delta F| \right\} \gg
\{ |H\delta \dot{F}|, |\delta \ddot{F}| \}\,.
\end{eqnarray}
The condition (\ref{percon2}) amounts to neglecting
the term $\delta \ddot{F}$ that leads to the oscillation
of $\delta F$.
This is the approximation used in scalar tensor models
in Refs.~\cite{Boi,review2,Tsuji}.
Later we explore the validity of such an approximation paying
particular attention to the conditions that should be satisfied.

Under the conditions (\ref{percon1}) and (\ref{percon2}),
Eq.~(\ref{delF}) gives
\begin{eqnarray}
\label{delRappro}
\delta R \simeq \frac{1}{F}
\frac{\delta \rho_m+3\dot{F} \dot{\delta}_m^{(v)}}
{1+3\xi}\,,
\end{eqnarray}
where
\begin{eqnarray}
\label{xidef}
\xi \equiv \frac{k^2}{a^2}\frac{f_{,RR}}{f_{,R}}
=\frac{k^2}{a^2R}m\,.
\end{eqnarray}
Using the approximation (\ref{percon2}) in Eq.~(\ref{delm}),
we obtain
\begin{eqnarray}
\label{metper1}
\ddot{\delta}_m^{(v)}+\left(2H+\frac{1}{1+3\xi}
\frac{\dot{F}}{2F} \right) \dot{\delta}_m^{(v)}
-4\pi G_{\rm eff}^{{\rm cosmo}} \rho_m
\delta_m^{(v)} \simeq 0\,,
\end{eqnarray}
where the ``cosmological'' effective gravitational constant
is given by
\begin{eqnarray}
\label{Gcosmo}
G_{\rm eff}^{{\rm cosmo}}
=\frac{G}{F} \left (\frac{1+4\xi}{1+3\xi} \right )\,.
\end{eqnarray}
Note that we have restored the bare gravitational constant $G$.

Introducing a physical wavelength $\ell=a/k$,
the parameter $\xi$ defined in Eq.~(\ref{xidef}) can be written as
\begin{eqnarray}
\xi=\frac{1}{\ell^2}\frac{f_{,RR}}{f_{,R}}
\simeq \frac13 \frac{1}{(M\ell)^2}\,,
\end{eqnarray}
where in the last approximate equality we have used the
approximate relation
$M^2 \simeq \frac{f_{,R}}{3f_{,RR}}$.

In the regimes $\xi \ll 1$, i.e., $(M\ell)^2 \gg 1$,
Eq.~(\ref{Gcosmo}) gives $G_{\rm eff}^{\rm cosmo} \simeq G/F$.
In this case $m \ll 1$ for sub-horizon modes ($k \gg aH$).
Thus the deviation from the $\Lambda$CDM model
is small, i.e., $|\dot{F}/HF| \ll 1$ in Eq.~(\ref{metper1}).
Hence the evolution of matter perturbations is similar to the one
in the standard GR case.
We recall again that this General Relativistic behaviour
can be realized  even for $\omega_{\rm BD}=0$
because of the presence of a potential
with a heavy scalar-field mass ($M^2 \gg k^2/a^2$).

In the regimes $\xi \gg 1$, i.e., $(M\ell)^2 \ll 1$,
Eq.~(\ref{Gcosmo}) gives
$G_{\rm eff}^{\rm cosmo} \simeq 4G/3F$. Thus in this case the
evolution of matter perturbations is different
from the one in the GR case because of the appearance of
the $4/3$ factor. If the mass of the Brans-Dicke scalar
field is light ($M^2 \ll k^2/a^2$), the cosmological
effective gravitational constant in Brans-Dicke theory
is given by $G_{\rm eff}^{\rm cosmo} \simeq \frac{G}{\phi}
\left ( \frac{4+2\omega_{\rm BD}}{3+2\omega_{\rm BD}} \right )$ \cite{Tsuji}.
Thus in the regime $\xi \gg 1$,
the $f(R)$ theories in the metric formalism
behave as the Brans-Dicke
theory (with $\omega_{\rm BD}=0$), with a light
scalar-field mass ($M^2 \ll k^2/a^2$).

%--------------------------------------------
\subsection{Longitudinal gauge ($\chi=0$)}
%--------------------------------------------

We shall also derive the approximate equations in the longitudinal
gauge ($\chi=0$) for sub-horizon modes satisfying
Eq.~(\ref{percon1}). We also use the notation $\alpha=\Phi$
and $\varphi=-\Psi$, which then gives the relation
$\Psi=\Phi+\delta F/F$ from Eq.~(\ref{met3}).
In addition to Eq.~(\ref{percon2}),
we impose the following conditions
\begin{eqnarray}
\label{loncon1}
|\dot{X}| \lesssim |HX|\,,\quad {\rm where} \quad
X=F, \dot{F}, \Phi, \Psi\,,
\end{eqnarray}
and
\begin{eqnarray}
\label{loncon2}
\left\{ \frac{k^2}{a^2}|\Phi|, \frac{k^2}{a^2}|\Psi|,
\frac{k^2}{a^2}|\delta F| \right\} \gg
\left\{ H^2 |B|, H^2 |\Phi|, H^2 |\Psi| \right\}\,.
\end{eqnarray}
If the deviation from the $\Lambda$CDM model is not
significant, the condition (\ref{loncon1}) is well satisfied.
The condition (\ref{loncon2}) is also satisfied for
sub-horizon modes given in Eq.~(\ref{percon1})
provided that $\Phi$, $\Psi$ and $B$
are of the same order.

Under these approximations we obtain, from
Eqs.~(\ref{mattere}), (\ref{met1}), (\ref{met4})
and (\ref{met5}), the following relations
\begin{eqnarray}
\label{delmapp}
& & \ddot{\delta}_m^{(\chi)}+
2H\dot{\delta}_m^{(\chi)}+\frac{k^2}{a^2}\Phi
\simeq 0\,,\\
\label{Phire}
& &\frac{k^2}{a^2} \Phi \simeq -\frac{1}{2F}
\left ( \frac{1+4\xi}{1+3\xi} \right )\delta \rho_m\,,\quad
\frac{k^2}{a^2} \Psi \simeq -\frac{1}{2F}
\left ( \frac{1+2\xi}{1+3\xi} \right ) \delta \rho_m\,,\quad
\delta F \simeq \frac{f_{,RR}}{f_{,R}} \left ( \frac{1}{1+3\xi} \right )
\delta \rho_m\,.
\end{eqnarray}
From Eq.~(\ref{met2}) the term $v$
is of the order of $H\Phi/\rho_m$ provided that the deviation
from the $\Lambda$CDM model is not significant.
Using Eq.~(\ref{Phire}) we find that the ratio
$3Hv/(\delta \rho_m/\rho_m)$ is of the order of
$(aH)^2/k^2$, which is much smaller than unity
for sub-horizon modes.
This gives $\delta_m^{(\chi)} \simeq \delta \rho_m/\rho_m$
in Eq.~(\ref{delchidef}).
{}From Eqs.~(\ref{delmapp}) and (\ref{Phire}) the matter
perturbation in the longitudinal gauge satisfies the following
approximate equation
\begin{eqnarray}
\label{metper2}
\ddot{\delta}_m^{(\chi)}+
2H \dot{\delta}_m^{(\chi)}
-\frac{\rho_m}{2F} \left (\frac{1+4\xi}{1+3\xi} \right )
\delta_m^{(\chi)} \simeq 0\,.
\end{eqnarray}

Compared to the comoving gauge the difference appears
only in the friction term.
Since viable $f(R)$ models satisfy the condition $|\dot{F}/HF| \ll 1$,
Eq.~(\ref{metper1}) reduces to Eq.~(\ref{metper2}).
We have also checked that in uniform density
gauge ($\delta \rho_m=0$) the perturbation
$\delta_m^{(\delta)}$ satisfies the same approximate
equation as Eq.~(\ref{metper2}).

Before ending this Subsection, we shall
introduce a number of parameters which can be useful below.
A useful parameter is the effective gravitational potential
\begin{eqnarray}
\label{Phieffdef}
\Phi_{\rm eff} \equiv (\Phi+\Psi)/2\,,
\end{eqnarray}
which characterises the deviation of light rays.
This is directly linked with the Integrated Sachs-Wolfe (ISW) effect
in the CMB \cite{Song,linear} and weak lensing of distant
galaxies \cite{Sapone,Tsuji}.
{}From Eq.~(\ref{Phire}) we can approximate this parameter by
\begin{eqnarray}
\label{Phieff}
\Phi_{\rm eff} \simeq -\frac{a^2}{2k^2}\frac{\rho_m}{F}
\delta_m^{(\chi)}\,.
\end{eqnarray}
We introduce an anisotropic parameter
\begin{eqnarray}
\label{eta}
\eta \equiv \frac{\Phi-\Psi}{\Psi} \simeq
\frac{2\xi}{1+2\xi}\,,
\end{eqnarray}
which behaves as $\eta \to 1$ for $\xi \gg 1$ and
$\eta \to 2\xi$ for $\xi \ll 1$.
We also define another variable
\begin{eqnarray}
\label{Sigma}
\Sigma \equiv q(1+\eta/2)\,,
\end{eqnarray}
where $q$ is defined via
$(k^2/a^2)\Psi=-(1/2)q\rho_m \delta_m^{(\chi)}$.
Using the above expressions $\Sigma$ can be
approximated by
\begin{eqnarray}
\label{Sigma2}
\Sigma \simeq 1/F\,.
\end{eqnarray}
Note that  $\Sigma$ is directly linked with $\Phi_{\rm eff}$.
The parameters $(\Sigma, \eta)$ can be especially important
in future survey of weak lensing \cite{Sapone,Tsuji}.

%------------------------------------------
\subsection{The appearance of scalarons}
%------------------------------------------

Among the approximations we have used in the previous two subsections,
the conditions (\ref{con3met}) and (\ref{percon2})
can be violated if an oscillating mode
(scalaron) dominates over the matter induced mode discussed above.
Let us clarify when the oscillating mode becomes important for
viable $f(R)$ models satisfying the conditions $m \ll 1$ and
$|\dot{F}/HF| \ll 1$.
For the sub-horizon modes, Eq.~(\ref{delF}) is approximately given by
\begin{eqnarray}
\delta \ddot{F}+3H\delta \dot{F}+\left(\frac{k^2}
{a^2} +M^2 \right)\delta F
 \simeq \frac13 \delta \rho_m\,.
\end{eqnarray}
The solution of this equation is the sum of the matter induced mode
$\delta F_{\rm ind}$ and the oscillatory scalaron
mode $\delta F_{\rm osc}$ satisfying
\begin{eqnarray}
\delta \ddot{F}_{\rm osc}+3H\delta \dot{F}_{\rm osc}+
\left(\frac{k^2}{a^2} +M^2 \right)\delta F_{\rm osc}=0\,.
\end{eqnarray}
Under the condition $\{M^2,k^2/a^2\} \gg H^2$ this equation reduces
to the form $(a^{3/2}\delta F_{\rm osc})^{\ddot{}}+
\omega^2 (a^{3/2}\delta F_{\rm osc}) \simeq 0$, where
$\omega=\sqrt{k^2/a^2+M^2}$.
In the adiabatic regime characterised by $|\dot{\omega}/\omega^2| \ll 1$
we obtain the following WKB solution
\begin{eqnarray}
\label{wkb}
\delta F_{\rm osc} \simeq ca^{-3/2}\frac{1}{\sqrt{2\omega}}
\cos \left( \int  \omega {\rm d}t \right)\,,
\end{eqnarray}
where $c$ is a constant.
Hence the solution of the perturbation $\delta R$ is expressed by
\begin{eqnarray}
\label{delrso}
\delta R \simeq \frac{1}{f_{,R}} \frac{1}{1+3\xi}
\delta \rho_m +c a^{-3/2}\frac{1}{f_{,RR}\sqrt{2\omega}}
\cos \left( \int  \omega {\rm d}t \right)\,.
\end{eqnarray}

For viable $f(R)$ models, the scale factor $a$ and the background Ricci scalar
$R^{(0)}$ evolve as $a \propto t^{2/3}$ and
$R^{(0)} \simeq 4/(3t^2)$ during the matter era.
Then the amplitude of $\delta R_{\rm osc}$ relative to $R^{(0)}$
has a time-dependence
\begin{eqnarray}
\frac{|\delta R_{\rm osc}|}{R^{(0)}} \propto
\frac{M^2t}{(k^2/a^2+M^2)^{1/4}}\,.
\end{eqnarray}
Let us consider the models $m(r)=C(-r-1)^p$ ($p>0$) for which
the mass $M$ evolves as $M \propto t^{-(p+1)}$
during the matter-dominated epoch.
When $\xi \ll 1$ and $\xi \gg 1$ we have
$|\delta R_{\rm osc}|/R^{(0)} \propto t^{-(3p+1)/2}$
and $|\delta R_{\rm osc}|/R^{(0)} \propto t^{-2(p+1/3)}$,
respectively.
Hence the amplitude of the oscillating mode
decreases faster than the background Ricci scalar.
Thus if the scalaron is over-produced in the early Universe
such that $|\delta R| >R^{(0)}$, the stability condition (\ref{con3met})
can be violated. This property persists in the radiation-dominated
epoch as well \cite{Starobinsky07,ShinjiNew}.
Thus in order to ensure the viability of the $f(R)$
theories of gravity in metric formalism, we need to esnure that
$|\delta R|$ is smaller than $R^{(0)}$
at the beginning of the radiation era. This can be achieved by
choosing the constant $c$ in Eq.~(\ref{wkb}) to be sufficiently
small which amounts to a fine tuning for these theories.
We note that this fine tuning concerns the stability of
these theories and is an additional constraint to those
usually imposed on the parameters of these theories by
observations.

Under the condition that the scalaron mode $\delta R_{\rm osc}$ is negligible
relative to the mater-induced mode $\delta R_{\rm ind}$, one can derive
the evolution for the matter perturbation $\delta_m$ as well as the effective
gravitational potential $\Phi_{\rm eff}$.
When $\xi \ll 1$ the evolutions of $\delta_m$ and $\Phi_{\rm eff}$
during the matter era are given by
\begin{eqnarray}
\delta_m \propto t^{2/3}\,,\quad
\Phi_{\rm eff}={\rm constant}\,.
\label{Phiso1}
\end{eqnarray}
Note that the ratio of the matter induced mode relative to the background
Ricci scalar evolves as $|\delta R_{\rm ind}|/R^{(0)} \propto t^{2/3}
\propto \delta_m$.
For the models that satisfy cosmological and local gravity constraints,
the Universe typically starts from the regime $\xi \ll 1$ and evolved
into the regime with $\xi \gg 1$ during the matter-dominated
epoch \cite{Starobinsky07,ShinjiNew}.
When $\xi \gg 1$, $\delta_m$ and $\Phi_{\rm eff}$ evolve as
\begin{eqnarray}
\delta_m \propto t^{(\sqrt{33}-1)/6}\,,\quad
\Phi_{\rm eff} \propto t^{(\sqrt{33}-5)/6}\,.
\label{Phiso2}
\end{eqnarray}
For the models $m(r)=C(-r-1)^p$, we have the time-dependence
$|\delta R_{\rm ind}|/R^{(0)} \propto t^{-2p+(\sqrt{33}-5)/6}$
in the regime $\xi \gg 1$.
This decreases more slowly relative to the ratio
$|\delta R_{\rm osc}|/R^{(0)} \propto t^{-2(p+1/3)}$,
so the scalaron mode tends to be unimportant with time.

In what follows we shall numerically solve
the exact perturbation equations in order to check the validity
of approximations used to
reach Eqs.~(\ref{metper1}), (\ref{metper2}) and (\ref{Phieff}).
We choose initial conditions such that scalaron mode
is suppressed relative to the matter induced mode, i.e.
$|\delta R_{\rm osc}^i| < |\delta R_{\rm ind}^i|$.
 We refer the reader to Ref.~[29] for a comprehensive and detailed
study of the scalaron mode. This study also gives the conditions
under which the scalaron mode dominates over the
matter induced mode at the initial stages.
%See Ref.~\cite{ShinjiNew} for the case in which the scalaron
%mode dominates over the matter induced mode at the initial stages.

%----------------------------------------------------------------
\subsection{Numerical study of the validity of approximations}
%----------------------------------------------------------------

In order to study the dynamics of matter perturbations in the
metric formalism we shall introduce the following
dimensionless variables \cite{AGPT}
\begin{eqnarray}
\label{xs}
x_{1} = -\frac{\dot{F}}{HF}\,,\quad
x_{2} =  -\frac{f}{6FH^{2}}\,, \quad
x_{3} = \frac{R}{6H^{2}}=\frac{\dot{H}}{H^{2}}+2\,.
\end{eqnarray}
In terms of these variables, the energy fraction
$\Omega_m$ of the pressureless matter and  the effective equation
of state $w_{\rm eff}$ are given by
\begin{equation}
\Omega_{m}\equiv\frac{\rho_{m}}{3FH^{2}}
=1-x_{1}-x_{2}-x_{3}\,,\quad
w_{\rm eff} \equiv -1-\frac23 \frac{\dot{H}}{H^2}
=-\frac13 (2x_3-1)\,.
\label{Omem}
\end{equation}
The evolution equations for the background
dynamics can then be expressed as \cite{AGPT}
\begin{eqnarray}
\label{x1}
x_1' & = & -1-x_{3}-3x_{2}+x_{1}^{2}-x_{1}x_{3}\,,\label{N1}\\
x_2' & = & \frac{x_{1}x_{3}}{m}-x_{2}(2x_{3}-4-x_{1})\,,\label{N2}\\
\label{x3}
x_3' & = & -\frac{x_{1}x_{3}}{m}-2x_{3}(x_{3}-2)\,,\label{N3}
\end{eqnarray}
where a prime denotes a derivative with respect to the number of
e-folding $N=\log\,(a)$.
For a later use we also introduce the variable $x_4 \equiv aH$,
which satisfies
\begin{eqnarray}
\label{x4}
x_4'=(x_3-1)x_4\,.
\end{eqnarray}

The matter epoch corresponds to the critical point
\begin{eqnarray}
P_{M}:~(x_{1},x_{2},x_{3})=\left(\frac{3m}{1+m},
-\frac{1+4m}{2(1+m)^{2}},\frac{1+4m}{2(1+m)}\right)\,,
~~
w_{{\rm eff}}=-\frac{m}{1+m}\,,
~~
\Omega_{m}=1-\frac{m(7+10m)}{2(1+m)^{2}}\,.
\label{matterepo}
\end{eqnarray}
Since $m$ needs to be much smaller than unity
during the matter era,  we have
$P_M: (x_1,x_2,x_3) \simeq (0,-1/2,1/2)$
for viable $f(R)$ models.
We shall consider a case in which the evolution proceeds
from the matter point $P_M$ to the de-Sitter point given by:
\begin{eqnarray}
P_{A} :~(x_{1},x_{2},x_{3})=(0,-1,2)\,,\quad
w_{{\rm eff}}=-1\,, \quad
\Omega_m=0\,,
\end{eqnarray}
which can easily be shown to lie
on the line $r=-2$ and is stable for $0 <m \le 1$.

%---------------------------------
\subsubsection{Comoving gauge}
%---------------------------------

In the comoving gauge, the perturbation Eqs.~(\ref{delm}) and
(\ref{delF}) can be rewritten in terms of the above variables thus:
\begin{eqnarray}
\label{coeq1}
& & \delta_m^{(v)''}+\left( x_3-\frac12 x_1 \right)
 \delta_m^{(v)'}-\frac32 (1-x_1-x_2-x_3) \delta_m^{(v)}
 \nonumber \\
& & =\frac12 \Biggl[ \biggl( \frac{k^2}{x_4^2}-3+3x_1
+9x_2+3x_3\biggr) \delta \tilde{F}+3(-2x_1+x_3-1)
\delta \tilde{F}'+3\delta \tilde{F}'' \Biggr]\,, \\
\label{coeq2}
& & \delta \tilde{F}''+\left(1-2x_1+x_3 \right)
\delta \tilde{F}'
+\left[ \frac{k^2}{x_4^2}-x_3+\frac{2x_3}{m}
+1-x_1+3x_2 \right]\delta \tilde{F}
 \nonumber \\
& & = (1-x_1-x_2-x_3)\delta_m^{(v)} -x_1 \delta_m^{(v)'}\,,
\end{eqnarray}
where $\delta \tilde{F} \equiv \delta F/F$.
The exact evolution of the matter perturbation can be obtained
by solving these equations together with the background equations
(\ref{x1})-(\ref{x4}) for $x_1$, $x_2$, $x_3$ and $x_4$.
Meanwhile, the approximate equation (\ref{metper1}) can be expressed
in terms of these variables as
\begin{eqnarray}
\label{coeq3}
\delta_m^{(v)''}+\left[x_3-\frac{x_1}{2(1+3\xi)}\right]
\delta_m^{(v)'}-\frac32 (1-x_1-x_2-x_3)
\left (\frac{1+4\xi}{1+3\xi} \right ) \delta_m^{(v)}
\simeq 0\,,
\end{eqnarray}
where
\begin{eqnarray}
\xi=\frac{k^2}{(aH)^2} \frac{m}{6x_3}\,.
\end{eqnarray}

Let us consider the case in which the condition $M^2 \gg k^2/a^2$
(i.e., $\xi \ll 1$) is satisfied.
Since $M$ needs to be large during the matter-dominated epoch
to satisfy LGC, this condition holds
in viable $f(R)$ models at the beginning of the matter era
for the modes relevant to large scale
structure \cite{Starobinsky07,ShinjiNew}.
Then the term $2x_3/m$ dominates over
the term $k^2/x_4^2$ in Eq.~(\ref{coeq2}),
which gives $\delta \tilde{F} \sim m \delta _m^{(v)}$
under the neglect of scalarons.
Hence the right hand side of Eq.~(\ref{coeq1}) can be neglected
relative to the left hand side, which means that Eq.~(\ref{coeq1})
reduces to Eq.~(\ref{coeq3}).
The above argument shows that, in the regime $\xi \ll 1$,
Eq.~(\ref{coeq3}) can be valid even for
super-Hubble modes as long as the contribution
of the scalaron is unimportant.
In this regime the matter perturbations evolve as in the case of
standard GR, i.e. $\delta_m^{(v)} \propto t^{2/3}$.

%---------------------------------------------------
\begin{figure}[H]
\begin{center}
\includegraphics[width=3.4in,height=3.2in]{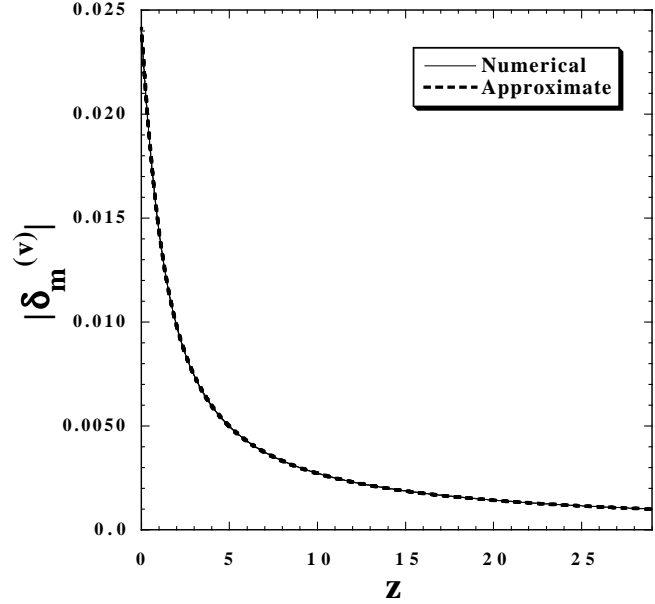}
\caption{\label{fig1}
The evolution of the matter perturbation $\delta_m^{(v)}$
in the comoving gauge for the model $m(r)=(-r-1)^3$
with the mode $k/a_0H_0=10$. Initial conditions were
chosen to be $x_1=0$, $x_2=-0.5000$, $x_3=0.5001$,
$\delta_m^{(v)}=10^{-3}$, $\delta_m^{(v)'}=10^{-3}$,
$\delta \tilde{F}=8.0 \times 10^{-15}$, $\delta \tilde{F}'=0$
and $k/a_iH_i=4.1$ at the redshift $z=28.9$.
The solid curve is obtained by solving the exact equations
(\ref{coeq1}) and (\ref{coeq2}) numerically,
whereas the dotted one is obtained
by solving the approximate equation (\ref{coeq3}).
}
\end{center}
\end{figure}
%----------------------------------------------------

The perturbations can enter the regime $M^2\ll k^2/a^2$ (i.e., $\xi \gg 1$)
before reaching the present epoch, depending on the mode $k$ and
on the evolution of $M$ \cite{Starobinsky07,ShinjiNew}.
In models $m(r)=(-r-1)^3$ this occurs for
the modes $k/a_0H_0>3.5$, where the subscript $0$
represent present values.
In the case $k/a_0H_0=300$, the redshift
at $k/a=M$ corresponds to $z_k=4.83$.
Since $M^2$ is always larger than $H^2$ in the past
because of the requirement $m \ll 1$,
the modes are inside the Hubble radius ($k^2/a^2H^2>1$)
after the perturbations enter the regime $M^2<k^2/a^2$.
Hence the approximation we used to reach Eq.~(\ref{coeq3})
is valid in this regime. In the regime $M^2<k^2/a^2$
the term $(k^2/x_4^2)\delta \tilde{F}$ in Eq.~(\ref{coeq2}),
balances the term $(1-x_1-x_2-x_3)\delta_m^{(v)}$, which gives rise
to an additional contribution on the right hand side of Eq.~(\ref{coeq1}).
This then leads to the approximate equation (\ref{coeq3})
with $\xi \gg 1$, which has a growing-mode solution
$\delta_m \propto t^{(\sqrt{33}-1)/6}$.

In Fig.~\ref{fig1} we plot the evolution of $\delta_m^{(v)}$
for the model $m(r)=(-r-1)^3$ with the mode $k/a_0H_0=10$.
Initial conditions are chosen so that the scalaron mode does
not dominate over the matter-induced mode.
In this case the transition from the regime $M^2>k^2/a^2$ to the region
$M^2<k^2/a^2$ occurs at
the redshift $z_k=1.62$.
We find that the approximate equation (\ref{coeq3})
shows an excellent agreement with
the results obtained by numerically solving
the exact equations (\ref{coeq1}) and (\ref{coeq2}).
The argument also holds for modes that are initially
outside the Hubble radius.
Thus the approximate equation (\ref{metper1})
is reliable to estimate the growth of matter perturbations
and the resulting matter power spectrum, provided that
the scalaron does not dominate in the early Universe.

%-----------------------------------------
\subsubsection{Longitudinal gauge}
%-----------------------------------------

In the longitudinal gauge the combination of
Eqs.~(\ref{met1})-(\ref{met5}) leads to the
following perturbation equations
\begin{eqnarray}
\label{numeeq1}
\hspace*{-1.0em}& &\Phi''+\left(2-\frac32 x_1+x_3\right)\Phi'
+(3x_2+3x_3)\Phi
=\frac32 x_2 \delta \tilde{F}-\left(\frac12 x_3+1\right)
\delta \tilde{F}'-\frac12 \delta \tilde{F}''\,, \\
\label{numeeq2}
\hspace*{-1.0em}& &\delta \tilde{F}''+(x_3+2)\delta \tilde{F}'+\left(
\frac43 \frac{k^2}{x_4^2}+3x_2+\frac{2x_3}{m}
\right) \delta \tilde{F}=
\left(6x_2+2x_3-\frac23 \frac{k^2}{x_4^2}
\right)\Phi-(3x_1+2)\Phi',\\
\label{numeeq3}
\hspace*{-1.0em}& &\delta_m^{(\chi)}=\frac{(2+3x_2-x_3+2x_3/m+k^2/x_4^2)
\delta \tilde{F}+(2+x_1+x_3)\delta \tilde{F}'+\delta \tilde{F}''
+(x_1-6x_2-2x_3)\Phi+(4x_1+2)\Phi'}
{1-x_1-x_2-x_3},  \nonumber \\
\\
\label{numeeq4}
\hspace*{-1.0em}& & \frac{\rho_m v}{FH}=2\Phi'+(2-x_1)\Phi+\delta \tilde{F}'
+(1+x_1)\delta \tilde{F}\,,
\end{eqnarray}
where we have used $\Psi=\Phi+\delta \tilde{F}$.
The effective potential defined in Eq.~(\ref{Phieffdef}) is given by
\begin{eqnarray}
\label{Phieffdef1}
\Phi_{\rm eff}=\Phi+\frac12 \delta \tilde{F}\,.
\end{eqnarray}

In order to understand the evolution of perturbations
at the initial stages of the matter era,
let us consider the regime  $\xi \ll 1$
without assuming the sub-horizon condition $k/(aH) \gg 1$.
We have in mind viable $f(R)$ models
with vanishingly small values of $m$ deep inside
the matter epoch.
Equation (\ref{numeeq2}) then becomes
\begin{eqnarray}
\label{delFsup}
\delta \tilde{F} \simeq -2m \left[ 1+\frac{k^2}
{3(aH)^2}\right]\Phi-2m\Phi'\,.
\end{eqnarray}
Note that under sub-horizon approximation
we have $\delta \tilde{F} \simeq -2mk^2\Phi/3(aH)^2$, which
agrees with Eq.~(\ref{Phire}).
Using Eq.~(\ref{delFsup}) we find that the right hand side
of Eq.~(\ref{numeeq1}) can be neglected relative to
the left hand side, thus giving the solution $\Phi={\rm constant}$
(together with a decaying mode proportional to $t^{-5/3}$).
{}From Eqs.~(\ref{Phieffdef}) and $(\ref{numeeq3})$
we obtain $\Phi_{\rm eff} \simeq \Phi$ and
\begin{eqnarray}
\label{deldif}
& &\delta_m^{(\chi)}
\simeq -\frac{2k^2}{3(aH)^2\Omega_m}\Phi_{\rm eff}\,,\\
\label{deldif2}
& &\delta_m^{(\chi)'} \simeq \delta_m^{(\chi)}\,.
\end{eqnarray}
Equation (\ref{deldif}) agrees with the expression (\ref{Phieff})
obtained under the sub-horizon approximation ($k/a \gg H$).
Since $\Phi_{\rm eff}$ is a constant, the matter perturbation
can be seen from Eq.~(\ref{deldif}) to evolve
as $\delta_m^{(\chi)} \propto a$.
This is consistent with the approximate
equation (\ref{metper2}), i.e.,
\begin{eqnarray}
\label{loneq1}
\delta_m^{(\chi)''}+x_3
\delta_m^{(\chi)'}-\frac32 (1-x_1-x_2-x_3)
\frac{1+4\xi}{1+3\xi}\delta_m^{(\chi)}
\simeq 0\,,
\end{eqnarray}
which has the growing mode solution
$\delta_m^{(\chi)}=\delta_m^{(\chi)'} \propto a$
in the regime $\xi \ll 1$.

We may ask why the above method reproduces the result
derived under the sub-horizon approximation, without employing the
approximation $k/a \gg H$.
In the regime $\xi \ll 1$ the perturbation $\delta \tilde{F}$ is suppressed
relative to $\Phi$ as given in Eq.~(\ref{delFsup}). This allows us to
neglect the right hand side of Eq.~(\ref{numeeq1}),
giving a constant $\Phi$.
This mimics the situation in General Relativity where
$\delta \tilde{F}=0$ and $\Phi={\rm constant}$ together
with Eq.~(\ref{deldif}), resulting in $\delta_m^{(\chi)} \propto a$.
Moreover, from Eq.~(\ref{numeeq4}), the quantity
$B=Hv+\Psi$ is well approximated by $B \simeq 5\Phi/3=\,$constant.
Hence the right hand side of Eq.~(\ref{mattere})
can be neglected even without assuming the sub-horizon approximation.
Thus using the relation (\ref{deldif}) we can
obtain Eq.~(\ref{metper2}) in the regime $\xi \ll 1$ without
assuming $k/a \gg H$.
The above approximation corresponds to the limit of large
$M$ ($M^2 \gg k^2/a^2$),
which gives rise to the evolution of perturbations close to
the case of General Relativity. In General Relativity
($\delta F=0$ and $\dot{F}=0$),
one has the exact equation
(\ref{deldif}) from Eqs.~(\ref{met1}) and (\ref{met2}).
Thus the perturbations in the large $M$ case ($\xi \ll 1$)
mimic those in General Relativity, apart from
the fact that the scalaron is present in the former
but not in the latter.

When $\xi \gg 1$ one has $k^2/a^2 \gg M^2 \gg H^2$,
which means that the sub-horizon type approximation
we used in the subsection B holds well in this regime.
This situation is similar to the case of the comoving gauge.
For the modes that start from the regime $M^2 \gg k^2/a^2$
and enter the regime $M^2 \ll k^2/a^2$ before the end of the
matter era, the evolution of perturbations changes
from the standard general relativistic form (\ref{Phiso1}) to
the non-standard form (\ref{Phiso2}).

\begin{figure}[H]
\begin{center}
\includegraphics[width=3.4in,height=3.4in]{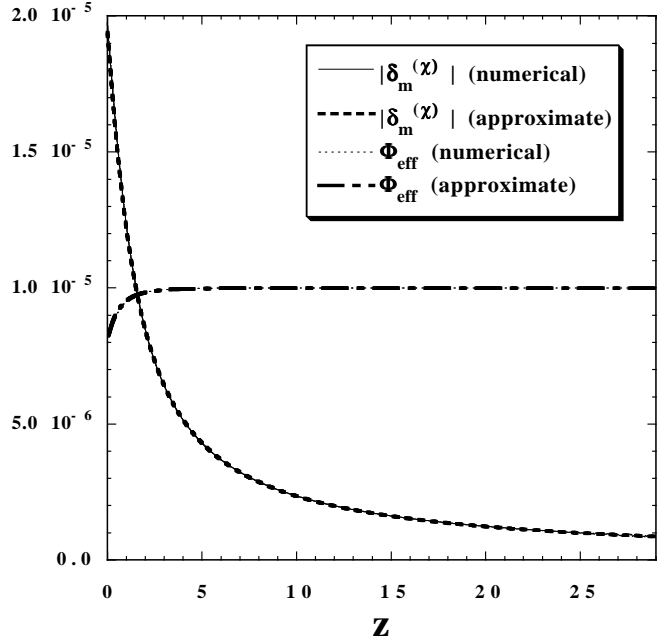}
\caption{\label{fig2}
The evolution of $\delta_m^{(\chi)}$ and $\Phi_{\rm eff}$
in the longitudinal gauge for the model
$m(r)=(-r-1)^3$ with a mode $k=a_0H_0$.
We numerically integrated Eqs.~(\ref{numeeq1}),
(\ref{numeeq2}) and (\ref{numeeq3}) with initial conditions
$\Phi_{\rm eff}=10^{-5}$, $\Phi_{\rm eff}'=0$ and
$k/a_iH_i=0.36$ and with $\delta_m^{(\chi)}$ and $\delta_m^{(\chi)'}$
satisfying Eqs.~(\ref{deldif}) and (\ref{deldif2}).
Initial conditions for the background quantities
were chosen to be the same as in Fig.~\ref{fig1}.
We also plot $\delta_m^{(\chi)}$ and $\Phi_{\rm eff}$
obtained by solving the approximate equations
(\ref{loneq1}) and (\ref{Phieff}).
The approximation is valid even when
the mode is initially outside the Hubble radius.}
\end{center}
\end{figure}

In Fig.~\ref{fig2} we plot the evolution of $\delta_m^{(\chi)}$
and $\Phi_{\rm eff}$ in the model $m(r)=(-r-1)^3$
for the mode $k=a_0H_0$ that lies outside the Hubble
radius at the start of integration ($z=28.9$).
Together with numerically integrating Eqs.~(\ref{numeeq1})-(\ref{numeeq3}),
we also solve the approximate equation (\ref{loneq1}) with $\Phi_{\rm eff}$
derived by (\ref{Phieff}).
{}From Fig.~\ref{fig2} we find that the approximate equations
agree well with the exact numerical results, even if the mode is
initially slightly outside the Hubble radius.
We caution, however, that for large-scale modes far outside
the Hubble radius the scalaron can be important.
In fact we have numerically checked that the oscillating mode appears
for such super-Hubble modes unless the coefficient of the scalaron
in Eq.~(\ref{delrso}) is fine-tuned to be small.
In Fig.~\ref{fig2} the growth of the gravitational potential is not seen
in the region $z<z_k$, since the transition redshift is small ($z_k=0.36$).
It can, however, be observed if we consider modes on smaller scales.

%---------------------------------------------------
%----------------------------------------------------

In summary, for viable $f(R)$ models that satisfy the cosmological
and local gravity constraints, the approximate
Eqs.~(\ref{loneq1}) and (\ref{Phieff})
are trustable even for the modes outside the Hubble radius,
as long as the scalaron is suppressed relative to
the matter-induced mode.

%---------------------------------------------------------
\subsection{Constraints on the model $m(r)=C(-r-1)^p$}
%---------------------------------------------------------

Finally we consider the current and future constraints
on models $m(r)=C(-r-1)^p$ with $0<C \le 1$.
At the background level, compatibility with the
SNIa observations could result in the divergence of
the equation of state of dark energy \cite{AT,ShinjiNew}.
Interestingly the redshift at which such a divergence may
occur could be of order unity.
However the current SNIa observations are
not yet sufficiently accurate to rule out such cases.
Some constraints on the model parameters can be obtained
from the present equation of state of dark energy,
but even models with $p=1.5$ and $C=1$
are allowed \cite{ShinjiNew}. Thus the background
does not provide strong constraints
on the model parameters.
However this situation can change in the future
when higher-redshift data will become available from the
observations of SNIa and Gamma Ray bursts.

\begin{figure}[H]
\begin{center}
\includegraphics[width=3.4in,height=3.4in]{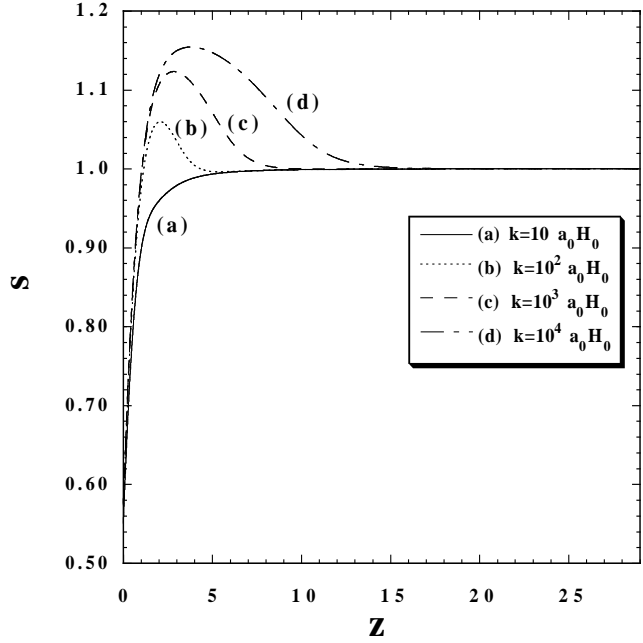}
\caption{\label{fig3}
The evolution of the growth rate $s=\delta_m'/\delta_m$
with respect to the redshift $z$ in the model $m(r)=(-r-1)^3$
with four different values of $k$.
Initial conditions were chosen as in Fig.~\ref{fig2}.
The transition redshift $z_k$ defined as the redshift
where $k/a=M$, becomes larger for smaller scales.
After the matter perturbation enters the region $z<z_k$
the growth rate begins to increase toward the value $s=1.186$,
but it starts to decrease once the Universe enters
the stage of accelerated expansion.}
\end{center}
\end{figure}

There are a number of additional
observational constraints on the growth rate of matter
perturbations. At the redshift $z\sim 3$,
McDonald {\it et al.} \cite{Mc} obtained the constraint
$\delta_m'/\delta_m=1.46 \pm 0.49$ from
the measurement of the matter power spectrum from the
Lyman-$\alpha$ forests.
Taking into account the more recent data reported by
Viel and Haehnelt \cite{Viel} in the redshift
range $2<z<4$, the maximum value of
the growth rate allowed by the current observations
is given by \cite{Diporto}
\begin{eqnarray}
\label{scon}
s \equiv \delta_m'/\delta_m  \lesssim 1.5\,.
\end{eqnarray}
The current data still have large error bars and some data
even allow the parameter range $-1<s<2$ \cite{Viel}.
However it is expected that in future observations the growth
rate will be constrained more severely.
{}From Eqs.~(\ref{Phiso1}) and (\ref{Phiso2})
we have $s=1$ for $M^2 \gg k^2/a^2$ and
$s=(\sqrt{33}-1)/4=1.186$ for $M^2 \ll k^2/a^2$.
In Fig.~\ref{fig3} we plot the evolution of the growth rate
for models $m(r)=(-r-1)^3$ for a number of different
values of $k$. The increase of $s$ from unity
implies that the perturbations enter the regime
$M^2 \ll k^2/a^2$.
For smaller scale modes this transition occurs earlier,
which leads to the larger maximum value of $s$.
The growth rate begins to decrease once the Universe
enters the late-time accelerated epoch.
As estimated analytically, the growth rate is bounded by
$s< 1.186$.
Hence the current observational constraint (\ref{scon}) is still
too weak to place constraints on $m(r)=C(-r-1)^p$
models.

%---------------------------------------------------
%----------------------------------------------------

However, these models exhibit peculiar features
in the matter power spectrum.
This is a consequence of the fact that
there is a transition redshift $z_k$
at which the growth rate begins to change from $s=1$ to $s=1.186$.
For the modes relevant to galaxy clusters
($k/a_0H_0={\cal O}(10^2)$),
this transition typically occurs during the matter-dominated
epoch (see Fig.~\ref{fig3}). Since the time $t_k$ at $z=z_k$
depends upon the modes $k$ ($t_k \propto k^{-3/(3p+1)}$),
this leads to the change in the slope of
the matter power spectrum.
The difference between the slopes of
the matter power spectrum determined
from galaxy surveys and the CMB spectrum,
on the scales $k/a_0H_0={\cal O}(10^2)$,
is given by \cite{Starobinsky07,ShinjiNew}
\begin{eqnarray}
\label{deln}
\Delta n \simeq \frac{\sqrt{33}-5}{3p+1}\,.
\end{eqnarray}
This analytic result agrees well with numerical
results except for models with $p \gg 1$ \cite{ShinjiNew}.
Observationally no significant differences have so far been found
between the two power spectra.
If we take the bound $\Delta n<0.05$ we obtain the
constraint $p \ge 5$.
To place further constraints on models,
a likelihood analysis is required which employs
the data from both the galaxy power spectrum and the CMB.
Such observational constraints, including the data from
the Lyman-$\alpha$ forest and Gamma Ray bursts,
are under consideration \cite{Acq}.

Numerically, we find that models $m(r)=(-r-1)^5$
have constraints on the present value of
the deviation parameter given by
$m(z=0) \lesssim 10^{-1}$.
Thus even though $m$ is constrained to be very small during the matter era,
a notable deviation from the $\Lambda$CDM
model can occur around the present epoch.

Finally the ISW effect in the CMB power spectrum is important
on large scale modes with $k/a_0H_0={\cal O}(1)$.
As can be seen from Fig.~\ref{fig2}, even models with $p=3$
and $C=1$ do not give rise to a significant amplification of the
gravitational potential. The models with $p \ge 2$ are
consistent with the low multipoles in the CMB data \cite{ShinjiNew}.
Thus this effect does not generally provide stronger
additional constraints.

%-----------------------------------------------------------
\section{Density perturbations in the Palatini formalism}
%-----------------------------------------------------------
\label{s5}

In this section we discuss the evolution of density perturbations
in the Palatini formalism and the resulting observational consequences.
The background equations are given by
\begin{eqnarray}
\label{Pabe1}
& & 6F \left( H+\frac{\dot{F}}{2F} \right)^2-f
=\rho_m\,, \\
\label{Pabe2}
& & FR-2f=-\rho_m\,,
\end{eqnarray}
where the Ricci scalar $R$ satisfies
the following relations
\begin{eqnarray}
\label{Req}
& & R=6(2H^2+\dot{H})+\frac{3}{F}
\left( \ddot{F}+3H\dot{F}-\frac{\dot{F}^2}{2F} \right)
\,,\\
\label{dotR}
& & \dot{R}=-\frac{3H \rho_m}{F-RF_{,R}}\,.
\end{eqnarray}

The matter perturbations satisfy Eqs.~(\ref{ma1d})
and (\ref{ma2d}) as in the case of the metric formalism.
Other perturbation equations
are \cite{Koivisto}
\begin{eqnarray}
\label{pa1}
& & -\frac{k^2}{a^2}\varphi+\left( H+\frac{\dot{F}}
{2F} \right)\kappa+\frac{1}{2F}
\left(\frac{3\dot{F}^2}{2F}+3H\dot{F} \right)\alpha
\nonumber \\
& &=\frac{1}{2F} \left[
\left( 3H^2-\frac{3\dot{F}^2}{4F^2}-\frac{R}{2}
+\frac{k^2}{a^2} \right) \delta F+
\left( \frac{3\dot{F}}{2F}+3H \right) \delta \dot{F}
-\delta \rho_m \right]\,,\\
\label{pa2}
& & H\alpha-\dot{\varphi}=\frac{1}{2F}
\left[ \delta \dot{F}- \left(H+\frac{3\dot{F}}{2F}
\right) \delta F-\dot{F} \alpha+\rho_m
v \right]\,, \\
\label{pa3}
& & \dot{\chi}+H \chi-\alpha-\varphi=
\frac{1}{F} ( \delta F-\dot{F}\chi)\,, \\
\label{pa4}
& & \dot{\kappa}+\left(2H+\frac{\dot{F}}{2F}\right)
\kappa+\left( 3\dot{H}+\frac{3\ddot{F}}{F}+
\frac{3H\dot{F}}{2F}-\frac{3\dot{F}^2}{F^2}
-\frac{k^2}{a^2} \right) \alpha+\frac32 \frac{\dot{F}}
{F}\dot{\alpha} \nonumber \\
& &=\frac{1}{2F} \left[ \delta \rho_m+
\left(6H^2+6\dot{H}+\frac{3\dot{F}^2}{F^2}-R+
\frac{k^2}{a^2}\right)\delta F+
\left( 3H-\frac{6\dot{F}}{F}
\right)\delta \dot{F}+3\delta \ddot{F}
 \right], \\
\label{pa5}
& & R\delta F-F \delta R=-\delta \rho_m\,.
\end{eqnarray}
Note that we corrected
several typos found in Ref.~\cite{Koivisto}.
Given the non-dynamical nature of Eq.~(\ref{pa5}),
it is clear that the scalaron mode does not exist
in the Palatini case.
This is associated with the fact that the Palatini formalism
corresponds to generalised Brans-Dicke theory (Eq.~(\ref{BD2})) with
$\omega_{\rm BD}=-3/2$.
The perturbation $\delta F$ is directly determined by the
matter perturbation $\delta \rho_m$, as
\begin{eqnarray}
\label{delFpa}
\delta F=\frac{F_{,R}}{F}
\frac{\delta \rho_m}{1-m}\,,
\end{eqnarray}
where $m$ is defined in Eq.~(\ref{mdef}).

Below we shall consider these perturbations in various gauges.

%----------------------------------------------
\subsection{Comoving gauge}
%----------------------------------------------

In the comoving gauge ($v=0$) one has
$\alpha=0$ and $\kappa=\dot{\delta}_m^{(v)}$.
Then from Eq.~(\ref{pa4}) we find
\begin{eqnarray}
\label{delmpa}
& &
\ddot{\delta}_m^{(v)}+\left(2H+\frac{\dot{F}}{2F}
\right) \dot{\delta}_m^{(v)} \nonumber \\
& &=\frac{1}{2F} \left[\left(6H^2+6\dot{H}+\frac{3\dot{F}^2}{F^2}-R+
\frac{k^2}{a^2}\right)\delta F+\left(3H-\frac{6\dot{F}}{F}\right)
\delta \dot{F}+3\delta \ddot{F}
+\delta \rho_m \right]\,.
\end{eqnarray}

Combining this with Eq.~(\ref{delFpa}) we obtain an exact
second-order equation for $\delta_m^{(v)}$, which can
be numerically solved (See Appendix A for details).
On the other hand, since we are
mostly interested in the evolution of modes on sub-horizon
scales, it makes sense to consider the approximate
equations similar to those considered in the metric case.

Using a sub-horizon type approximation, such that
only the terms containing $k^2/a^2$ and $\delta \rho_m$
are considered on the right hand side of Eq.~(\ref{delmpa}),
together with Eq.~(\ref{delFpa}), we obtain the
following approximate perturbation equation
\begin{eqnarray}
\label{delmpa2}
\ddot{\delta}_m^{(v)}+\left(2H+\frac{\dot{F}}{2F}
\right) \dot{\delta}_m^{(v)}-\frac{\rho_m}{2F}
\left(1+\frac{\xi}{1-m} \right)
\delta_m^{(v)} \simeq 0\,,
\end{eqnarray}
where $\xi$ is defined in Eq.~(\ref{xidef}).

Alternatively we may study the case in which the
deviation from the $\Lambda$CDM model is small, i.e.,
\begin{eqnarray}
\label{mcon}
|m| \ll 1\,,
\end{eqnarray}
as required from LGC (\ref{PalaLGC}).
The derivative terms such as $|m'|$ and $|m''|$
are also assumed to be much smaller than unity.
Using the fact that from Eqs.~(\ref{Pabe2}) and (\ref{delFpa})
the perturbation $\delta F$ in this case is of the order of
$m\delta_m^{(v)}$,
Eq.~(\ref{delmpa2}) can be obtained under condition (\ref{mcon})
without using the sub-horizon approximation.
Thus, if the deviation from the $\Lambda$CDM model is small,
the approximate equation (\ref{delmpa2}) is valid
even for the modes outside the Hubble radius.
This situation is similar to the case in the metric formalism.
In fact we have confirmed this property
by numerically solving the exact equation (\ref{delmpa})
and comparing it with the solutions of the
approximate equation (\ref{delmpa2}).

One can estimate the order of the term $R \delta F$
on the r.h.s. of Eq.~(\ref{pa4}) by using Eq.~(\ref{delFpa}), i.e.,
$R \delta F=m \delta \rho_m/(1-m)$.
This gives rise to the contribution of the order of
$(\rho_m/2F)m \delta_m^{(v)}$ in the third term
of Eq.~(\ref{delmpa2}), which is negligible under
the condition (\ref{mcon}).
As long as we neglect this contribution,  we should
approximate $\xi/(1-m) \simeq \xi$
in the third term of Eq.~(\ref{delmpa2}).
In the following, we implicitly assume this when we write
the term $(1-m)$ in the denominator.

In the limit $\xi=\frac{k^2}{a^2R}m \ll 1$,
Eq.~(\ref{delmpa2}) agrees with
Eq.~(\ref{metper1}) in the metric formalism.
However a significant difference appears in the regime
$\xi \gg 1$. In that case there is a
strong amplification of the matter perturbation
in the Palatini case because of the growth of
the term $\xi$ in Eq.~(\ref{delmpa2}).
We shall estimate this growth rate for
a number of concrete models in Subsection C
below.

%----------------------------------
\subsection{Longitudinal gauge}
%----------------------------------

We next consider the Longitudinal gauge ($\chi=0$),
and as in the metric case we use the notation
$\alpha=\Phi$ and $\varphi=-\Psi$.
Under the sub-horizon type approximation
used in the comoving case above,
the evolution equation reduces to Eq.~(\ref{delmapp})
obtained in the metric case.
Using Eqs.~(\ref{pa1}) and (\ref{pa3}) together with
Eq.~(\ref{delFpa}), these approximations also give
\begin{eqnarray}
\frac{k^2}{a^2}\Phi \simeq -\frac{1}{2F}
\left( 1+\frac{\xi}{1-m} \right) \delta \rho_m\,,
\quad
\frac{k^2}{a^2}\Psi \simeq -\frac{1}{2F}
\left( 1-\frac{\xi}{1-m} \right) \delta \rho_m\,.
\end{eqnarray}
Hence the matter perturbations satisfy the following
approximate equation
\begin{eqnarray}
\label{delmpa3}
\ddot{\delta}_m^{(\chi)}+
2H \dot{\delta}_m^{(\chi)}
-\frac{\rho_m}{2F}
\left(1+\frac{\xi}{1-m} \right)
\delta_m^{(\chi)} \simeq 0\,.
\end{eqnarray}
The effective gravitational potential $\Phi_{\rm eff}$
defined in Eq.~(\ref{Phieffdef}) satisfies
\begin{eqnarray}
\label{Poipala}
\Phi_{\rm eff} \simeq -\frac{a^2}{2k^2}
\frac{\rho_m}{F} \delta_m^{(\chi)}\,,
\end{eqnarray}
which is the same as in the metric case.
Similarly the parameters $\eta$ and $\Sigma$ defined
in Eqs.~(\ref{eta}) and (\ref{Sigma})
in this case become
\begin{eqnarray}
\eta \simeq \frac{2\xi}{1-m-\xi}\,,\quad
\Sigma \simeq \frac{1}{F}\,.
\end{eqnarray}
We note that while the expression for $\eta$
is different from that in the
metric case, $\Sigma$ remains the same.

The above approximate equations (\ref{delmpa3}) and
(\ref{Poipala}) are valid under the conditions
(\ref{mcon}) and $\xi \ll 1$
even without the sub-horizon approximation.
In fact the argument is similar to the metric case
in which Eqs.~(\ref{metper2}) and (\ref{Phieff})
reduce to the corresponding GR equations for $\xi \ll 1$.

We also note that in regimes $\xi \gg 1$
the perturbation modes are inside the
Hubble radius, which shows that the sub-horizon
approximation above is still valid.
Thus, as long as the condition (\ref{mcon}) is
satisfied, we can safely use Eqs.~(\ref{delmpa3})
and (\ref{Poipala}) even for super-Hubble modes.
Also since in the Palatini case
the perturbation $\delta R$ is sourced by
the matter induced mode only, we do not need to worry about
the dominance of the scalaron oscillations for super-Hubble modes,
unlike in the metric case.

In Ref.~\cite{Koivisto} the equation for matter perturbations
was derived in the uniform density gauge ($\delta \rho_m=0$).
This is an exact equation as in the comoving gauge
(see also Ref.~\cite{Uddin}).
We shall show in Appendix B that similar to other
gauges Eq.~(\ref{mcon}) is sufficient to reduce the
exact equation to the approximate one.
We shall also show that the approximate equation is the
same as the one in the longitudinal gauge (\ref{delmpa3}).

Since the evolution of matter perturbations hardly
depends on the gauge chosen, we shall in what follows
denote the matter perturbations simply by $\delta_m$.

%%%%%%%%%%%%%%%%%%%%%%%%%%%%%%%%%%%%%%%%%%%%%%%%%%%%%%%%%%%%%%%%%%%%%%%%%%%%%%
\subsection{Analytic estimate for the growth of perturbations}
%%%%%%%%%%%%%%%%%%%%%%%%%%%%%%%%%%%%%%%%%%%%%%%%%%%%%%%%%%%%%%%%%%%%%%%%%%%

As we mentioned above the evolution of perturbations in the regime
$\xi \ll 1$ is similar to the standard GR case where
$\delta_m \propto t^{2/3}$, $s=\delta_m'/\delta_m=1$
and $\Phi \propto$\,constant.
In this subsection we shall estimate the growth rate of
perturbations after the system enters the regime $\xi>1$.
We shall consider models with $|m| \ll 1$,
for consistency with LGC (\ref{PalaLGC}).

During the matter era in which the Ricci scalar evolves as
$R \propto t^{-2}$ the parameter $\xi$ is given by
$\xi= \pm ma/m_ka_k$, where the subscript ``$k$''
denotes the values when the system crosses $\xi=1$.
Here we note that the plus sign corresponds to a positive
$m$, whereas the negative sign to a negative $m$.
As we showed in Sec.~II,  the latter case can be allowed
unlike the metric case.
Under the condition $|m| \ll 1$, the matter perturbation
(\ref{delmpa3}) satisfies the following equation
\begin{eqnarray}
\label{delmpaap}
\delta_m''+\frac12 \delta_m'-\frac32
\left(1\pm \frac{m}{m_k}e^{N-N_k}\right)
\delta_m \simeq 0\,.
\end{eqnarray}

Lets us consider the case in which the evolution of
the parameter $m$ is given by
\begin{eqnarray}
m \propto t^{2p}\,,
\end{eqnarray}
where $p$ is a constant.
The values of $p$ in several different $f(R)$ models are given by
\begin{itemize}
\item (i) $f(R)=\alpha R^{1+m}-\Lambda$ (constant $m$):~~$p=0$\,,
\item (ii) $f(R)=R-\lambda R_c \left(\frac{R}{R_c}\right)^{\beta}$:~~
$p=1-\beta$~~for~~$R \gg R_c$\,,
\item (iii) $f(R)=R-\lambda R_c \frac{(R/R_c)^{2n}}
{(R/R_c)^{2n}+1}$:~~$p=2n+1$~~for~~$R \gg R_c$\,,
\item (iv) $f(R)=R-\lambda R_c \left[ 1-
\left( 1+\frac{R^2}{R_c^2} \right)^{-n} \right]$:~~$p=2n+1$
~~for~~$R \gg R_c$\,.
\end{itemize}
With the above choice of $m$, Eq.~(\ref{delmpaap}) reduces to
\begin{eqnarray}
\label{delmpaap2}
\delta_m''+\frac12 \delta_m'-\frac32
\left[1\pm e^{(3p+1)(N-N_k)}\right]
\delta_m \simeq 0\,.
\end{eqnarray}
For the positive sign in Eq.~(\ref{delmpaap2}), i.e., for $m>0$,
the solution of Eq.~(\ref{delmpaap2}) can be written in terms of
a linear combination of Bessel functions $J_\nu$ and $Y_\nu$:
\begin{eqnarray}
\delta_m=e^{-(N-N_k)/4} \left[ c_1 J_{\nu}
\left(ix\right)+c_2
Y_\nu \left( ix \right) \right]\,,
\end{eqnarray}
where $c_1, c_2$ are constants and
\begin{eqnarray}
\label{xnu}
x=\frac{\sqrt{6}e^{(3p+1)(N-N_k)/2}}{3p+1}\,,
\quad \nu=\frac{5}{6p+2}\,.
\end{eqnarray}
For the negative sign in Eq.~(\ref{delmpaap2}),  i.e., for $m<0$,
the solution of Eq.~(\ref{delmpaap2}) is given by
\begin{eqnarray}
\label{delmsolex2}
\delta_m=e^{-(N-N_k)/4} \left[ c_1 J_{\nu}
\left(x\right)+c_2
Y_\nu \left(x \right) \right]\,,
\end{eqnarray}
where $x$ and $\nu$ are as given in (\ref{xnu}).

In the following we shall discuss the positive and negative
$m$ cases in turn.

%------------------------------
\subsubsection{$m>0$}
%------------------------------

As an example, we consider the constant $m$ models ($p=0$).
In this case, the Bessel function $J_\nu (ix)$ has a growing mode
solution $J_{5/2} (ix) \propto I_{5/2}(x) \propto e^x/\sqrt{x}$
for $x \gg 1$ (here $I_{5/2}(x)$ is a modified Bessel
function with $x=\sqrt{6}e^{(N-N_k)/2}$).
Then, in the regimes $\xi \gg 1$,
the evolution of the matter perturbations
and its growth rate are given by
\begin{eqnarray}
\label{delmcons}
\delta_m \propto \exp(\sqrt{6} e^{(N-N_k)/2})\,,\quad
s=\frac{\delta_m'}{\delta_m}=\frac{\sqrt{6}}{2}
e^{(N-N_k)/2}\,,
\end{eqnarray}
where we have used $\sqrt{6}e^{(N-N_k)/2} \gg (N-N_k)/2$.
Thus the growth rate $s$ of the matter perturbations
increases very rapidly.
Also from Eq.~(\ref{Poipala}), in the regimes  $\xi \gg 1$, the
effective gravitational potential grows (double) exponentially as
\begin{eqnarray}
\label{Phicons}
\Phi_{\rm eff} \propto \exp (e^{\sqrt{6}(N-N_k)/2})\,,
\end{eqnarray}
which leads to a strong and observable ISW effect.

Similarly, in models with $p \neq 0$, one can estimate
the evolution of perturbations in the regime $\xi \gg 1$:
\begin{eqnarray}
\label{delses}
\delta_m \propto \Phi_{\rm eff} \propto
\exp \left( \frac{\sqrt{6}
e^{(3p+1)(N-N_k)/2}}{3p+1}\right)\,,\quad
s=\frac{\sqrt{6}}{2} e^{(3p+1)(N-N_k)/2}\,.
\end{eqnarray}
This shows that for models with $p>0$ the growth
rate increases faster than in the constant $m$ models.
When $p<-1/3$ the above instability can be avoided,
but in that case $m$ increases towards the past.
Thus unless the present value of $m$ is negligibly small,
the condition $|m| \ll 1$ required for LGC can be violated
during the matter era.
Thus these models are hardly distinguishable from the
$\Lambda$CDM model in the present Universe.
In what follows we shall concentrate on the positive
$p$ case.

We note that for these positive $m$ models
the violent growth of $\delta_m$ and $\Phi_{\rm eff}$
comes to an end after the Universe enters
the accelerated stage, since $\xi$ and $\Omega_m$
begin to decrease.

%----------------------
\subsubsection{$m<0$}
%----------------------

When $m$ is negative, the Bessel functions
in the regime $|\xi| \gg 1$ behave as
$J_\nu (x) \sim \sqrt{2/\pi x} \cos
[x-(2\nu+1)\pi/4]$ and
$Y_\nu (x) \sim \sqrt{2/\pi x} \sin
[x-(2\nu+1)\pi/4]$ respectively.
Thus the solution of Eq.~(\ref{delmsolex2})
in this asymptotic region becomes
\begin{eqnarray}
\delta_m \simeq C e^{-(3p+2)(N-N_k)/4}
\,\cos (x+\theta)\,, \quad
(|\xi| \gg 1)\,,
\end{eqnarray}
where $C$ and $\theta$ are constants.
Using this solution, we obtain
\begin{eqnarray}
\delta_m' &\simeq& -\frac14 (3p+2) \delta_m
-\frac{\sqrt{6}}{2}C e^{3p(N-N_k)/4}
\sin (x+\theta)\,, \\
\label{delms2}
s &\simeq& -\frac14 (3p+2) -\frac{3p+1}{2}
x \tan (x+\theta)\,,
\end{eqnarray}
which are also valid in the regimes $|\xi| \gg 1$.
When $p>0$, $\delta_m$
exhibits damped oscillations whereas $|\delta_m'|$
increases in time with the oscillations.
The averaged value of the growth rate $s$ is given by
$\bar{s}=-(3p+2)/4$, but it shows a divergence every time
$x$ changes by $\pi$.

If the Universe crosses the critical point $|\xi|=1$ around the end
of the matter era, it does not necessarily reach the regime $|\xi| \gg 1$.
In such cases one can not fully use the above approximate solutions.
We shall show later that, in some cases, the Universe can enter the
accelerated stage without oscillations in $\delta_m$ up to
the present epoch.
The oscillations in $\delta_m$ can be seen
as we choose larger values of $|m|$ and $k$.
The frequency of oscillations tends to grow
for larger values of $p$.
The models that enter the regimes $|\xi| \gg 1$
generally have difficulty in being consistent with observations,
since they lead to largely negative values of $s$ as
given by Eq.~(\ref{delms2}).

%--------------------------------------------------------------------
\subsubsection{Constraints on $|m|$ from the requirement $|\xi|<1$}
%--------------------------------------------------------------------

The $f(R)$ models can be consistent with observations
if the Universe does not enter the regime $|\xi|>1$
until the end of the matter-dominated epoch.
One can estimate the ratio of the comoving Hubble radius
$(aH)^{-1}$ during the matter era to its present value thus:
\begin{eqnarray}
\label{aeq}
\frac{a_0H_0}{aH} \simeq c\left(\frac{a}{a_0}
\right)^{1/2} = c(1+z)^{-1/2}\,,
\end{eqnarray}
where $c=1$ in the absence of the dark energy
dominated epoch.
The presence of a dark energy era leads to
a change in the value of $c$.
Numerically this factor is around $c=1.7$-$1.9$.
Using the relation $R \simeq 3H^2$ that holds during the
matter era for $|m| \ll 1$, we find that $|\xi|$
crosses 1 at a critical redshift
\begin{eqnarray}
\label{zces}
z_c \approx |m| \left( \frac{k}{a_0H_0}\right)^2-1\,.
\end{eqnarray}
If $z_c$ is smaller than the order of unity, the
Universe does not enter the regime $|\xi|>1$
during the matter dominated epoch.
This gives the following constraint:
\begin{eqnarray}
\label{mdemand}
|m(z)| \lesssim \left(\frac{a_0H_0}{k}\right)^2\,,~~~
{\rm for}~~~z>{\cal O}(1)\,.
\end{eqnarray}

The matter power spectrum, in the linear regime,
has been observed for the scales
$0.02 h\,{\rm Mpc}^{-1} \lesssim k \lesssim 0.2 h\,{\rm Mpc}^{-1}$.
Non-linear effects can be important for smaller scales with
$k>0.2h\,{\rm Mpc}^{-1}$.
Taking the value $k=0.2h\,{\rm Mpc}^{-1} \simeq 600 a_0H_0$, below
which linear perturbation theory is valid, we obtain the constraint
$|m(z)| \lesssim 3 \times 10^{-6}$ during the matter era.

Of course this is a rough estimate and the actual constraints
on $m(z)$ depend upon the particular models considered.
For example, even if $|\xi|$ crosses 1 during the matter era,
the models can be consistent with observations provided
that $|\xi|$ does not grow rapidly after the crossing.
Whether or not $\xi$ reaches the regime
$|\xi| \gg 1$ depends
on the particular models chosen.
Hence to place constraints on $m$,
we need a detailed analysis for each particular model.
In the next subsection we shall provide a numerical
investigation of a number of $f(R)$ models presented
above and place constraints on present values of $m$
as well as model parameters.

%%%%%%%%%%%%%%%%%%%%%%%%%%%%%%%%%%%%%%%%%%%%%%%%%%%%%%%%%%%%%%%%%%%%%%%%%%%%%
\subsection{Constraints on model parameters}
%%%%%%%%%%%%%%%%%%%%%%%%%%%%%%%%%%%%%%%%%%%%%%%%%%%%%%%%%%%%%%%%%%%%%%%%%%%%%

In this subsection we shall employ the information provided
by the growth of the matter density perturbations to place constraints
on the parameters of the $f(R)$ models presented in
subsection C above. We do this by numerically solving the exact perturbation
(\ref{CLdelmpa}) together with the background equations (\ref{Pabe1})
and (\ref{Pabe2}). See Appendix C for equations written in a form
convenient for numerical integration.

%----------------------------------------------------------------
\subsubsection{Constant $m$ models: $f(R)=\alpha R^{1+m}-\Lambda$}
%----------------------------------------------------------------

Compared to other models considered here,
the growth of $|\xi|$ is rather mild in the
constant $m$ models, being of the form $|\xi| \propto a =e^N$.
Thus, in order for $|\xi|$ to grow from $0.1$ to $10$, one
would require an increase in the number of e-foldings by 4.6.

\begin{figure}[H]
\includegraphics[width=3.4in,height=3.4in]{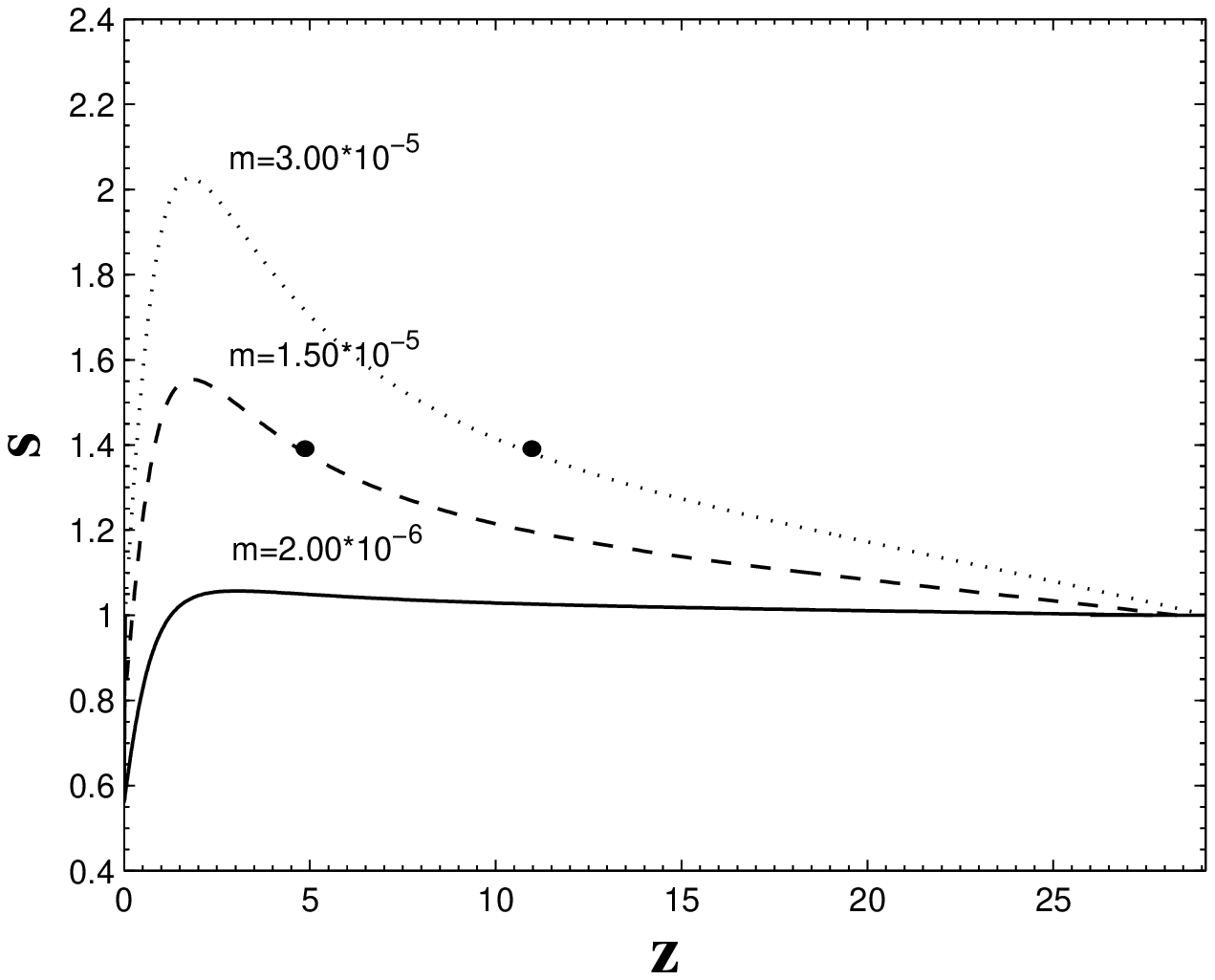}
\includegraphics[width=3.4in,height=3.4in]{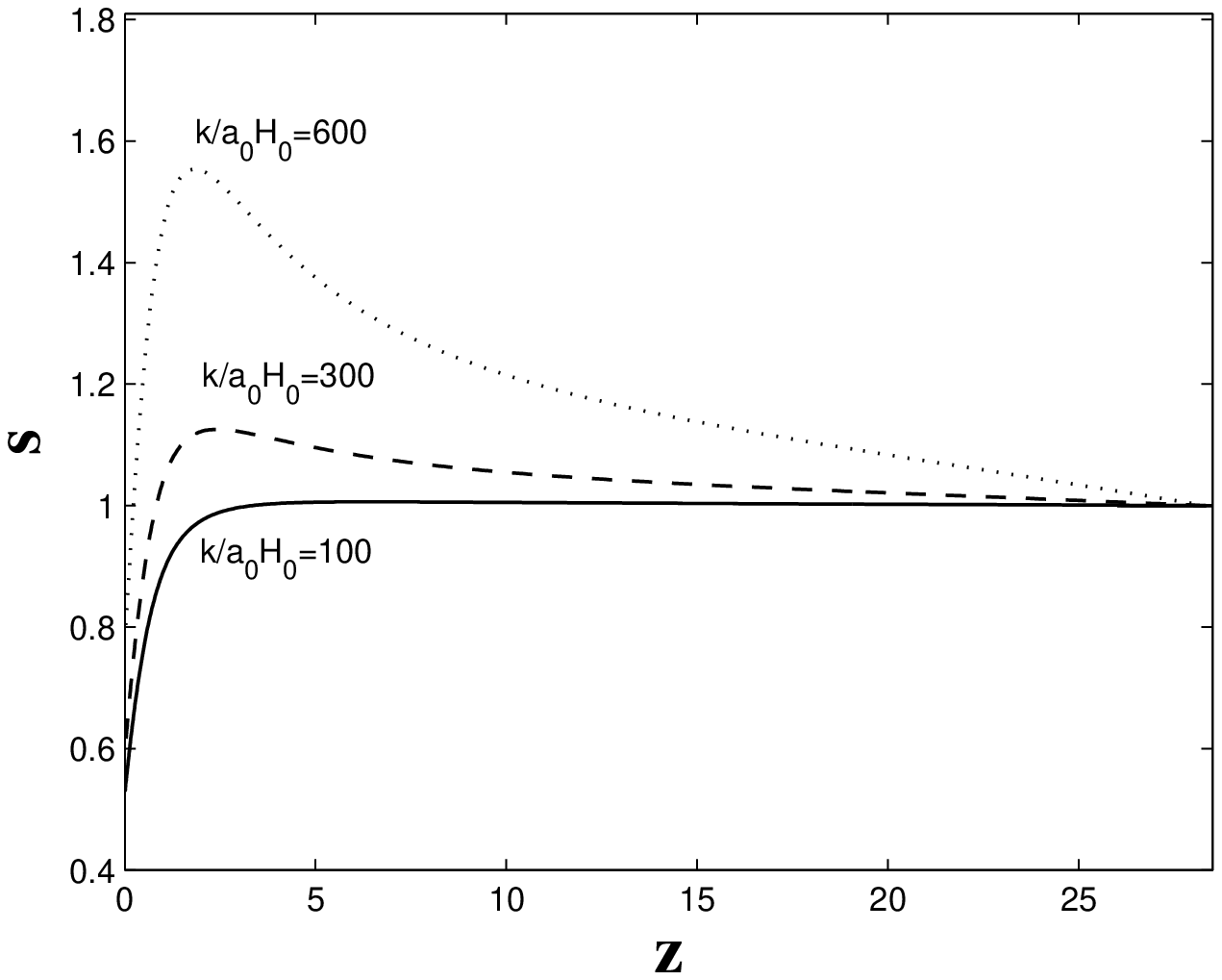}
\caption{\label{fig4}
The evolution of perturbations for the model:
$f(R)=\alpha R^{1+m}-\Lambda$ with positive values of $m$.
In the left panel we show the growth rate $s=\delta'_m/\delta_m$
versus the redshift $z$ for the mode $k/a_0H_0=600$
with three different values of $m$.
The black dots represent the points at which
$\xi$ crosses 1.
The right panel depicts the evolution of $s$
for $m=1.5 \times 10^{-5}$
with three different values of $k$.}
\end{figure}

We shall first consider the positive $m$ case.
In the left panel of Fig.~\ref{fig4} we plot the evolution of
growth rate $s=\delta_m'/\delta_m$ for the mode $k=600 a_0H_0$
for several values of $m$.
For $m=3 \times 10^{-5}$ we numerically obtain
$z_c \sim 11$, denoted by a black dot in Fig.~\ref{fig4}.
This almost agrees with the analytical estimate (\ref{zces})
which gives $z_c \approx 10$.
In the regimes $\xi \ll 1$ the evolution of matter perturbations
is given by $\delta_m=\delta_m' \propto a$, which results in
$s \simeq 1$.
The growth rate $s$ begins to move away from unity
as $\xi$ becomes of order 0.1, and then
continues to grow before the Universe enters
the stage of accelerated expansion. For this model
we find  $s_{\rm max} \sim 2.06$ and
$\xi_{\rm max} \sim 3.13$,
which shows that the model does not enter the regime
$\xi \gg 1$ where the evolution of perturbations
is described by Eqs.~(\ref{delmcons})
and (\ref{Phicons}).

For a model with $m=1.5 \times 10^{-5}$,
the critical redshift occurs at around $z_c\sim 5$ with $s\sim 1.4$.
The maximum value of the growth rate
is $s_{\rm max} \sim 1.57$,  which corresponds
to the marginal case satisfying the criterion (\ref{scon}).
For a model with $m=2.0 \times 10^{-6}$, the evolution of
perturbations is not much different from the
general relativistic case.

%---------------------------------------------------
%----------------------------------------------------

To show the variation of the growth rate
as a function of scale, we depict in the
right panel of Fig.~\ref{fig4} the evolution of $s$
for the model $m=1.5 \times 10^{-5}$ for three different
values of $k$. As can be seen, the maximum
value of the growth rate $s$
decreases as $k$ is decreased (i.e. the scales become larger).
In particular, for the mode $k=100a_0H_0$ (corresponding to
$k=0.33h$ Mpc$^{-1}$), the evolution of perturbations
exhibits no difference compared to the corresponding
evolution in the general relativistic case.
Hence the matter power spectrum is enhanced
on small scales ($k=0.1h$-$0.2h$ Mpc$^{-1}$),
while the spectrum remains similar to the
standard general relativistic case on larger scales
($k=0.02h$-$0.04h$ Mpc$^{-1}$).
This results in different spectral indices
on different scales.
Placing more precise constraints on $m$, would require
performing a likelihood analysis using the data from
the matter power spectrum.
However, in order to obtain an order of magnitude for
the maximum value of $m$, it is sufficient
to use the criterion (\ref{scon})
for the mode $k=600a_0H_0$.
For the constant $m$ models we find the constraint to be
$m \lesssim 10^{-5}$.

\begin{figure}[H]
\includegraphics[width=3.4in,height=3.4in]{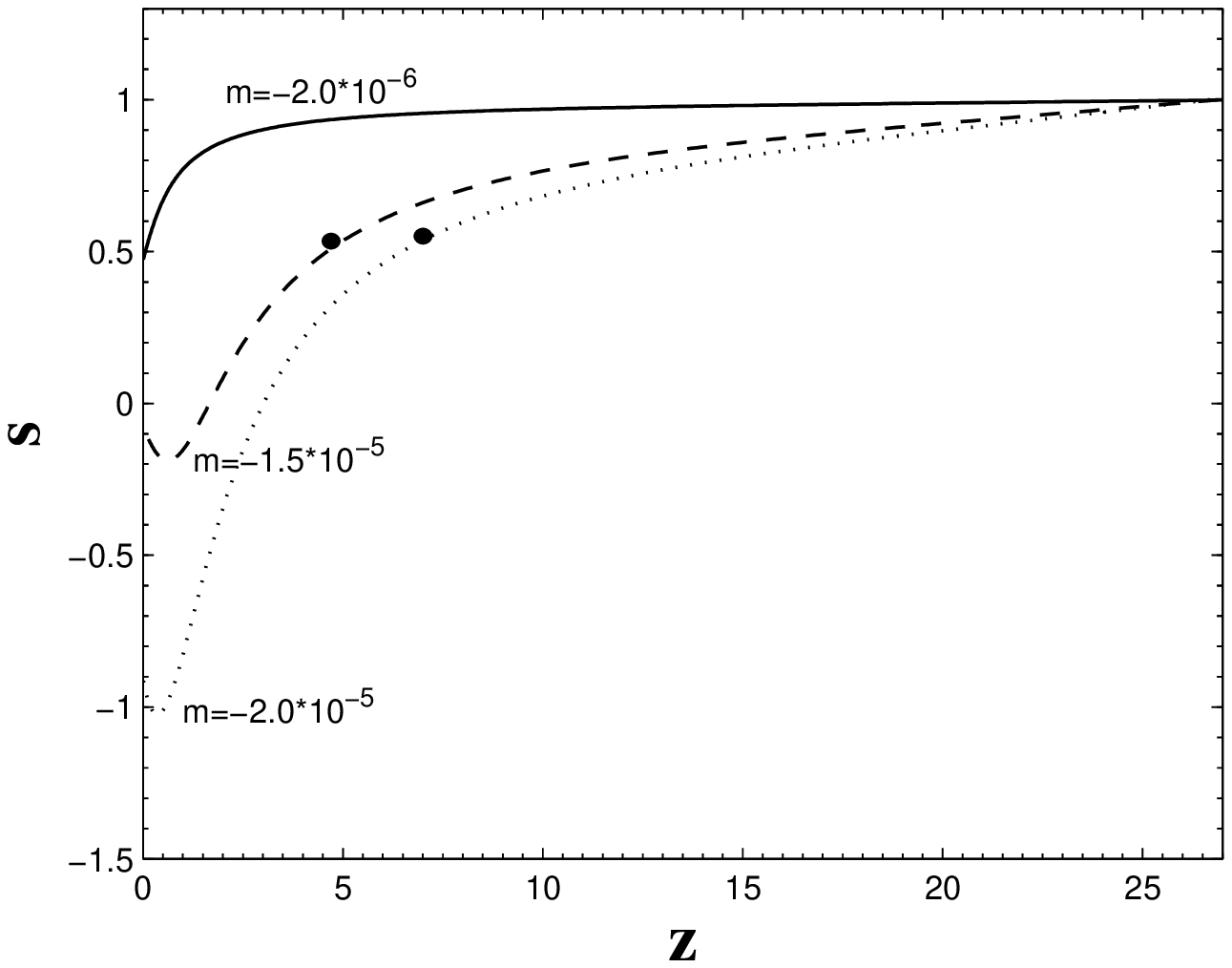}
\includegraphics[width=3.4in,height=3.4in]{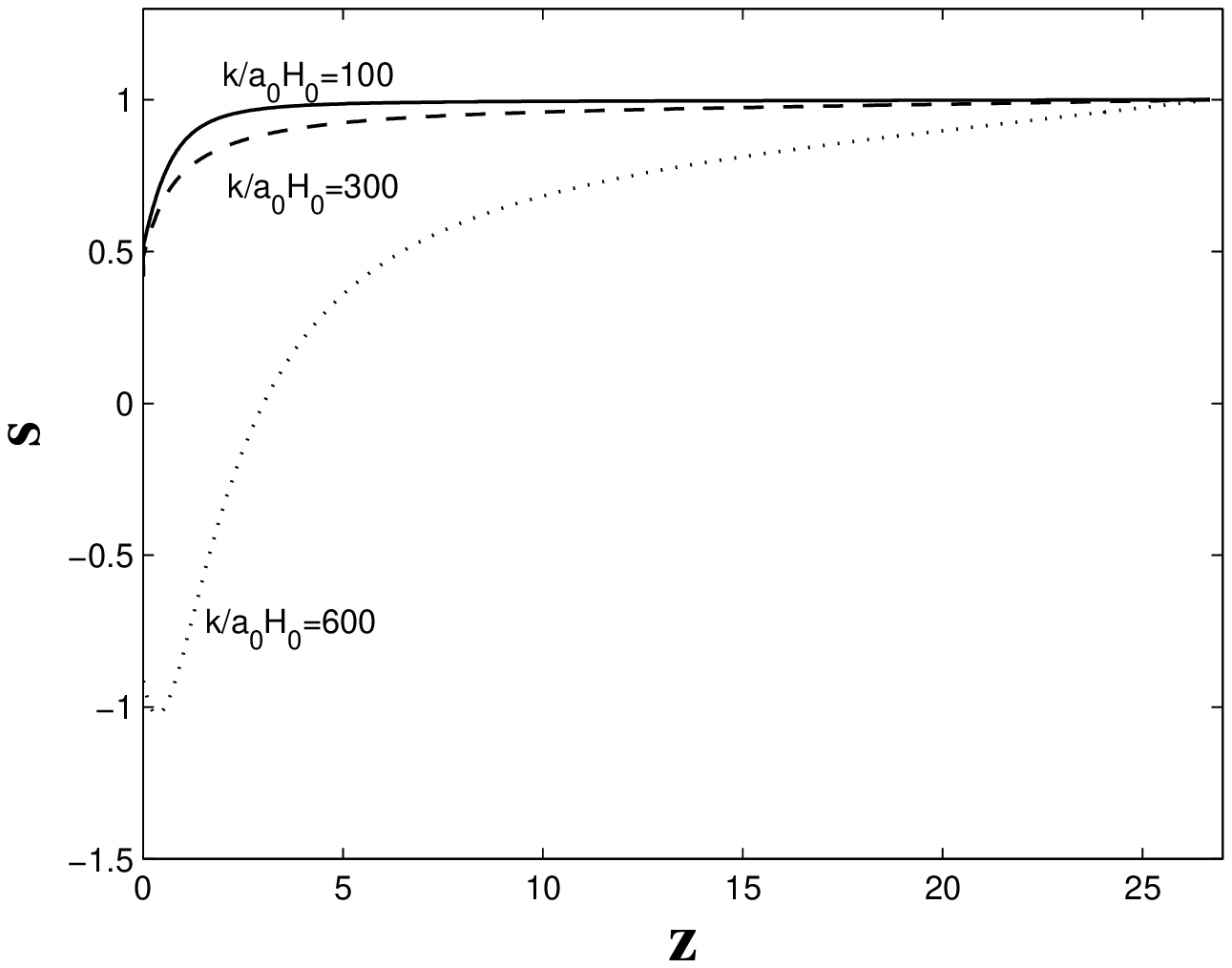}
\caption{\label{fig5}
The evolution of perturbations for the model:
$f(R)=\alpha R^{1+m}-\Lambda$ for negative $m$.
In the left panel we show the growth rate $s=\delta'_m/\delta_m$
versus the redshift $z$ for the mode $k/a_0H_0=600$
with three different values of $m$.
The black dots represent the points at which the quantity
$|\xi|$ crosses 1.
The right panel depicts the evolution of $s$
for $m=-2.0 \times 10^{-5}$
with three different values of $k$.}
\end{figure}

When $m$ is negative, the growth rate $s$ decreases
unlike the positive $m$ case.
In the left panel of Fig.~\ref{fig5} we plot
the evolution of $s$ for three different negative values of $m$
for the mode $k/a_0H_0=600$.
As can be seen
$s$ tends to decrease more rapidly with increasing $|m|$.
If $m=-2.0 \times 10^{-5}$ the present value of
$s$ becomes very small ($s< -1$).
As we see in the right panel of Fig.~\ref{fig5},
when $m=-2.0 \times 10^{-5}$, there is a significant
fall in the values of $s$ between
$k/a_0H_0=300$ and $k/a_0H_0=600$.
This can lead to large differences in the
spectral indices of the matter power spectrum for
small and large scale modes.
{}From the above argument $|m|$ should be smaller
than the order of $10^{-5}$,
which has an upper bound similar to
the positive $m$ case.

When $m=-2.0 \times 10^{-5}$ the Universe crosses
the point $|\xi|=1$ at the redshift $z_c \sim 7.4$,
but the increase of $|\xi|$ for $z<z_c$ is mild.
Moreover the quantity $|\xi|$ begins to decrease
after the Universe enters the accelerated stage.
Numerically we obtain the value $\xi \sim -0.77$ at present
($z=0$). Thus the system does not reach the regime
$|\xi| \gg 1$, and hence not a single period of oscillation
occurs by the present epoch. However,
for larger values of $|m|$, we have
numerically checked that the oscillations
of $\delta_m$ indeed occur.

%---------------------------------------------------
%----------------------------------------------------

We also recall that the growth of effective gravitational
potential $\Phi_{\rm eff}$ leads to an ISW effect
in the CMB spectrum on large scales
($k/a_0H_0 \sim$a few).
However, when $|m| \sim 10^{-5}$, $\Phi_{\rm eff}$
does not grow for these modes. As a result
the ISW effect does not provide stronger
constraints $m$ than those
provided by the matter power spectrum.

%----------------------------------------------------
\subsubsection{$f(R)=R-\lambda R_c (R/R_c)^{\beta}$}
%----------------------------------------------------

The observational constraints on the parameters of
this model were studied in Ref.~\cite{Koivisto}.
[Note that $R_c$ is not very much different from the present
value of the cosmological Ricci scalar $R_0$.]
Here, we shall obtain constraints on the parameter $m$
which for this model is given by
\begin{eqnarray}
m =\frac{\lambda \beta (1-\beta) (R/R_c)^{\beta-1}}
{1-\lambda \beta (R/R_c)^{\beta-1}}\,,
\end{eqnarray}
and make a comparison between our results.
The late-time de-Sitter point ($R=R_1$) is obtained
from the constraint equation $FR-2f=0$, to give
$(R_1/R_c)^{1-\beta}=\lambda (2-\beta)$.
Thus at this de-Sitter point
the variable $m$ satisfies
\begin{eqnarray}
\label{mbeta}
m(R_1)=\beta/2\,.
\end{eqnarray}
For $\beta<1$, the parameter $m$ in the
regime $R \gg R_c$ is given by
\begin{eqnarray}
\label{mevo}
m \simeq \lambda \beta (1-\beta)
(R/R_c)^{\beta-1} \propto t^{2(1-\beta)}\,,
\end{eqnarray}
which decreases towards the past.

If $\beta\,(<1)$ is of the order of unity,
the quantity $m$ is too large to satisfy the requirement
(\ref{mdemand}) for the mode $k=600a_0H_0$
during the matter era
(recall that from Eq.~(\ref{mbeta})
the present value of $m$ is of the order of $\beta$).
This is basically associated with the
fact that, in the regimes $R \gg R_c$,
the model gives a linear relation between $m$ and $r$
[$m=C(-r-1)$].
Thus we need the condition $|\beta| \ll 1$
in order to be compatible
with the criterion (\ref{mdemand}).

\begin{figure}[H]
\includegraphics[width=3.4in,height=3.4in]{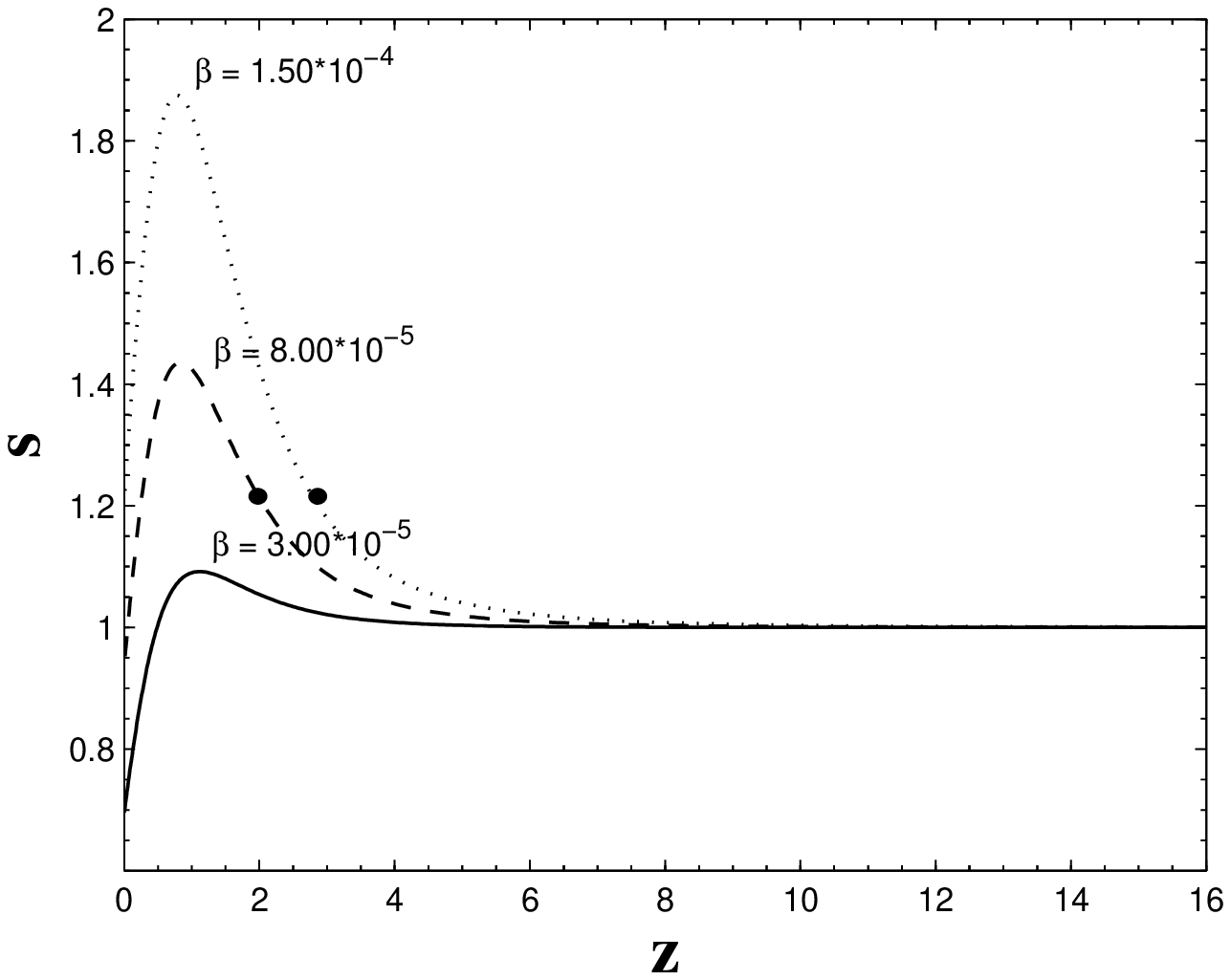}
\includegraphics[width=3.4in,height=3.4in]{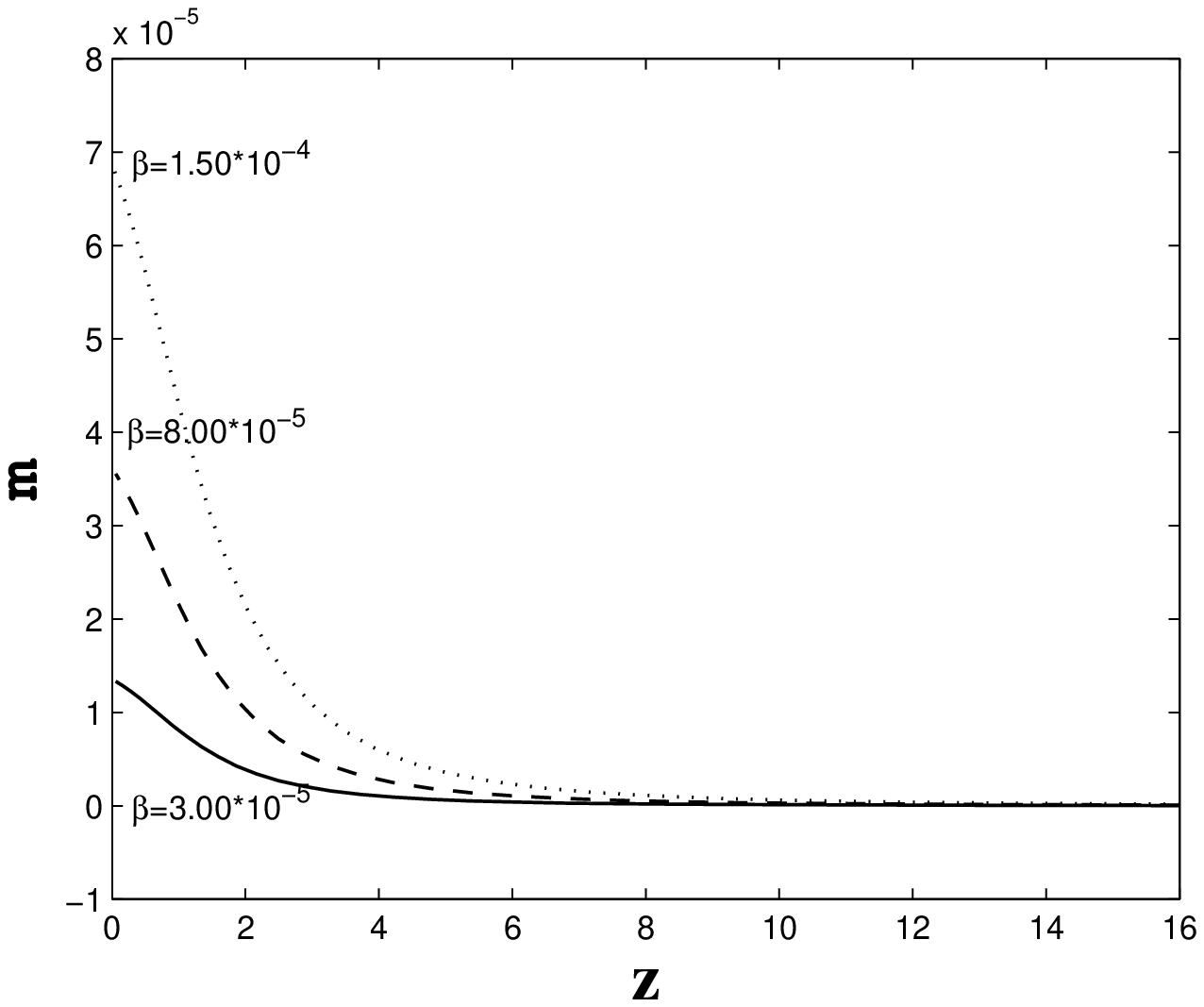}
\caption{\label{fig6}
The evolution of perturbations for the model
$f(R)=R-\lambda R_c (R/R_c)^{\beta}$ with
positive $\beta$ and $\lambda=1$.
The left hand panel depicts $s=\delta'_m/\delta_m$
versus the redshift $z$ for the mode $k/a_0H_0=600$
with three different values of $\beta$.
The right hand panel shows the evolution of
$m$ with respect to $z$ for $k/a_0H_0=600$.
From the requirement (\ref{scon})
we obtain the constraint $\beta<8.2 \times 10^{-5}$.
}
\end{figure}

To determine the changes in the behaviour of this model
as a function of $\beta$, we considered three distinct
values of $\beta$ and calculated the corresponding
growth rate $s$ and the parameter $m$ in each case.
Our results are summarised in Fig.~\ref{fig6}.
The left hand panel shows the
evolution of the growth rate $s$ for $\lambda=1$
and $k=600a_0H_0$
with the three different values of $\beta$.
For $\beta=1.5 \times 10^{-4}$ the
present value of the parameter $m$ is around
$m(z_0) \sim 6.7 \times 10^{-5}$, which is close to
the value of $m$ at the de-Sitter point
($m(R_1)=7.5 \times 10^{-5}$).
We also find that the parameter $\xi$ crosses 1
at a redshift $z_c \sim 3$
with $m(z_c) \sim 1.2 \times 10^{-5}$.

%---------------------------------------------------
%----------------------------------------------------

Furthermore, we find that the growth
rate $s$ is larger for these models than
in the case of constant $m$ models.
This is due to the fact that $\xi$
in this case evolves
faster, as $\xi \propto t^{2(4/3-\beta)}$.
The maximum growth rate reached for
$\beta=1.5 \times 10^{-4}$
corresponds to $s_{\rm max}\sim 1.88$ with $\xi \sim 4$.
As expected,  models with smaller values of $\beta$
possess growth rates which are more compatible with
observational constraints.
Employing the criterion
(\ref{scon}) for the mode $k=600a_0H_0$, we find the constraint
$\beta<8.2 \times 10^{-5}$.
This is slightly larger than the constraint
$\beta<3.0 \times 10^{-5}$ obtained in \cite{Koivisto06} from
the likelihood analysis of the SDSS data \cite{comment}.
In the left panel of Fig.~\ref{fig6} we also
consider this case in order to find the
corresponding evolution of $s$.
The maximum value of the growth rate in this case
is found to be $s_{\rm max} \sim 1.095$,
which indicates that the constraint (\ref{scon}) is rather weak.
Nevertheless, the criterion (\ref{scon}) is certainly
sufficient in order to extract the order of magnitude
of the bound on $\beta$.

In the right panel of Fig.~\ref{fig6}
we plot the evolution of
the parameter $m$ for the case with $\lambda=1$ and
$k=600a_0H_0$, for three values of $\beta$.
As can be seen $m$ increases from the past to
the present. Using the criterion (\ref{scon}) we obtain
the bound $m(z=0)<3.5 \times 10^{-5}$.
If we adopt the severer criterion $s<1.1$, the constraint
becomes $m(z=0)<1.3 \times 10^{-5}$.
Thus the deviation from the $\Lambda$CDM model
is constrained to be small ($m(z=0) \lesssim 10^{-5}$).

We also studied the effects of changing the parameter $\lambda$
in the action on the bounds on $\beta$. We considered two cases with
$\lambda=10$ and $\lambda=100$.
We found that these changes in $\lambda$ have negligible effects
on the constraints imposed on $\beta$ and $m(z=0)$, compared to that
obtained from the case with $\lambda=1$.
The reason for this lack of sensitivity is that a
change in parameter $\lambda$ is compensated by
corresponding changes to the values of $R_i, a_0$ and $H_0$.

\begin{figure}[H]
\includegraphics[width=3.4in,height=3.4in]{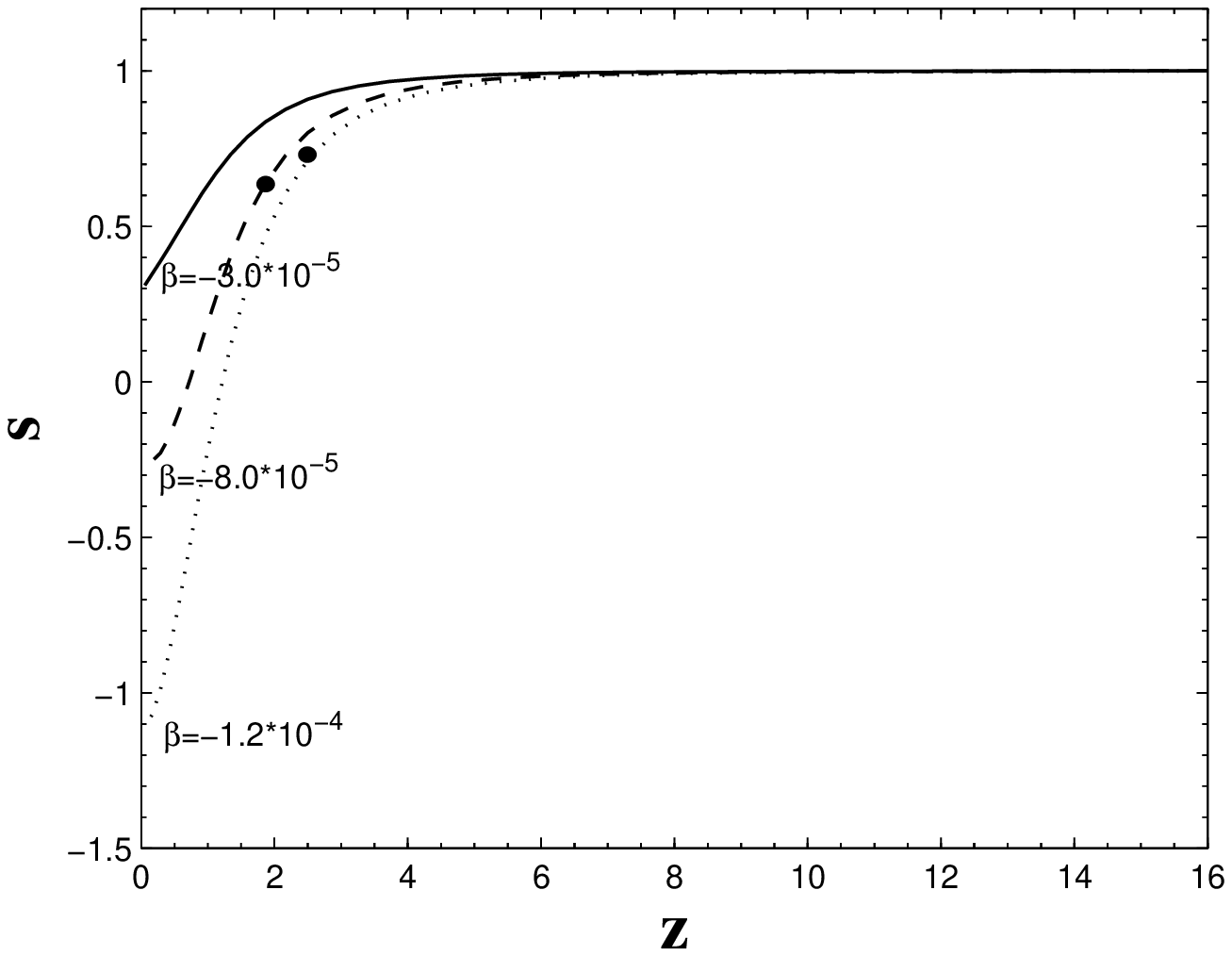}
\includegraphics[width=3.4in,height=3.4in]{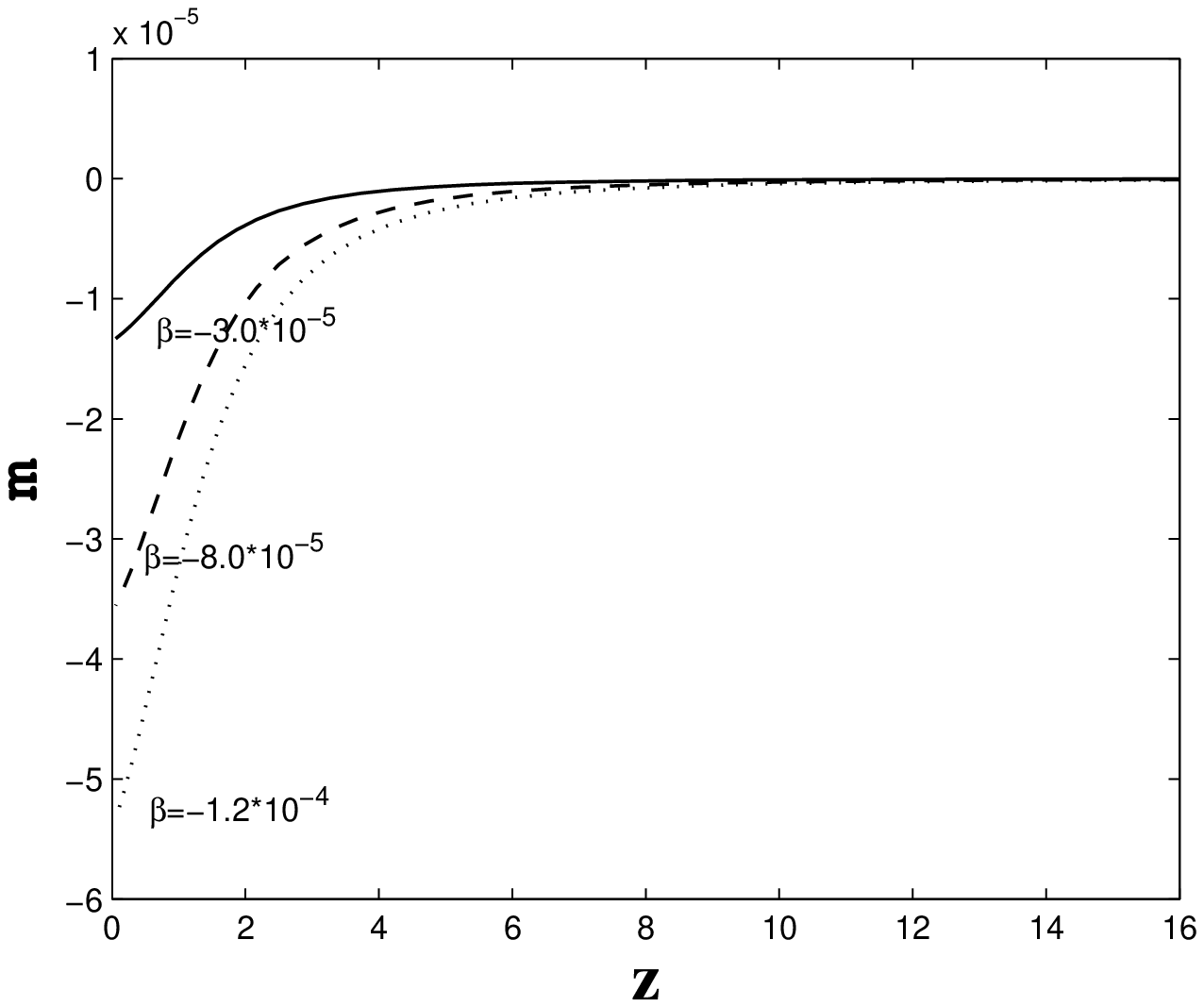}
\caption{\label{fig7}
The evolution of perturbations for the model
$f(R)=R-\lambda R_c (R/R_c)^{\beta}$ with
negative $\beta$ and $\lambda=1$.
The left hand panel depicts $s=\delta'_m/\delta_m$
versus the redshift $z$ for the mode $k/a_0H_0=600$
with three different values of $\beta$.
The right hand panel shows the evolution of the quantity
$m$ with respect to $z$ for $k/a_0H_0=600$.
If we use the criterion $s(z=0)>-1$, we obtain
the constraint $\beta > -1.2 \times 10^{-4}$.
}
\end{figure}

When $\beta<0$ the parameter $m$ is negative from
Eq.~(\ref{mevo}).
In the left panel of Fig.~\ref{fig7} we plot the evolution
of $s$ for three different values of $\beta$ with
$k/a_0H_0=600$.
We find that the present values of $s$ become smaller
than $-1$ for $|\beta| \gtrsim 1.2 \times 10^{-4}$,
in which case $|m(z=0)|$ is smaller than
the order of $5.3 \times 10^{-5}$
(see the right panel of Fig.~\ref{fig7}).
Thus if we use the criterion $s(z=0) \gtrsim -1$
for the validity of the models, the upper bounds
of $|\beta|$ and $|m(z=0)|$ are similar to those
in the positive $\beta$ case.

%---------------------------------------------------
%----------------------------------------------------

%----------------------------------------------------
\subsubsection{$f(R) = R-\lambda R_{c}
[1-(1+R^2/R_c^2)^{-n}]$}
%----------------------------------------------------

Finally we consider the above model (where $n>0$) recently discussed by
Starobinksy \cite{Starobinsky07}.
The parameter $m$ for this model is given by
\begin{eqnarray}
m =\frac{2n\lambda x(1+x^2)^{-n-2}[(2n+1)x^2-1]}
{1-2n\lambda x(1+x^2)^{-n-1}}\,,~~~
{\rm where}~~~
x\equiv R/R_c,
\end{eqnarray}
and the de-Sitter point at $R=R_1$ corresponds to
\begin{eqnarray}
\label{lamratio}
\lambda=\frac{x_1(1+x_1^2)^{n+1}}
{2[(1+x_1^2)^{n+1}-1-(n+1)x_1^2}\,,~~~
{\rm where}~~~
x_1 \equiv R_1/R_c\,.
\end{eqnarray}
Once we fix the value of $\lambda$, $x_1$ is known accordingly.
In the regime $R \gg R_c$ the parameter $m$ behaves as
\begin{eqnarray}
\label{mevo2}
m \simeq 2n(2n+1)\lambda (R_c/R)^{2n+1}
\propto t^{2(2n+1)}\,.
\end{eqnarray}
Because of the presence of a larger power of $(R_c/R)$
in the expression for $m$ in this case,
$m$ decreases more rapidly towards the past compared to
the model $f(R)=R-\lambda R_c (R/R_c)^{\beta}$
discussed above.
For the mode $k=600a_0H_0$, the bound (\ref{mdemand}) implies
that $m$ has to be smaller than the order
of $10^{-6}$-$10^{-5}$ by the end of the matter-dominated epoch,
in order to ensure that the model does not enter
the regime with $\xi>1$.

\begin{figure}[H]
\includegraphics[width=3.4in,height=3.4in]{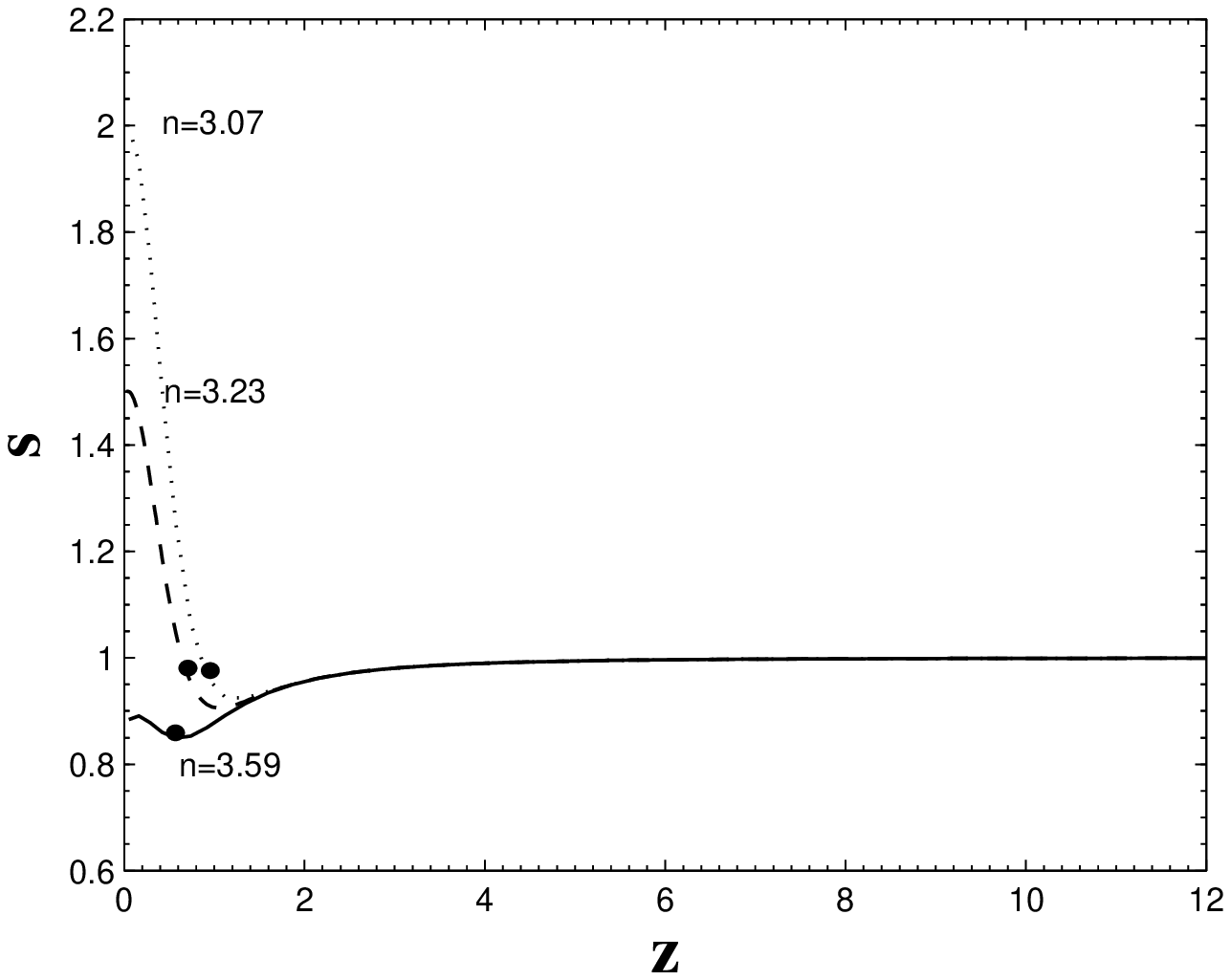}
\includegraphics[width=3.4in,height=3.4in]{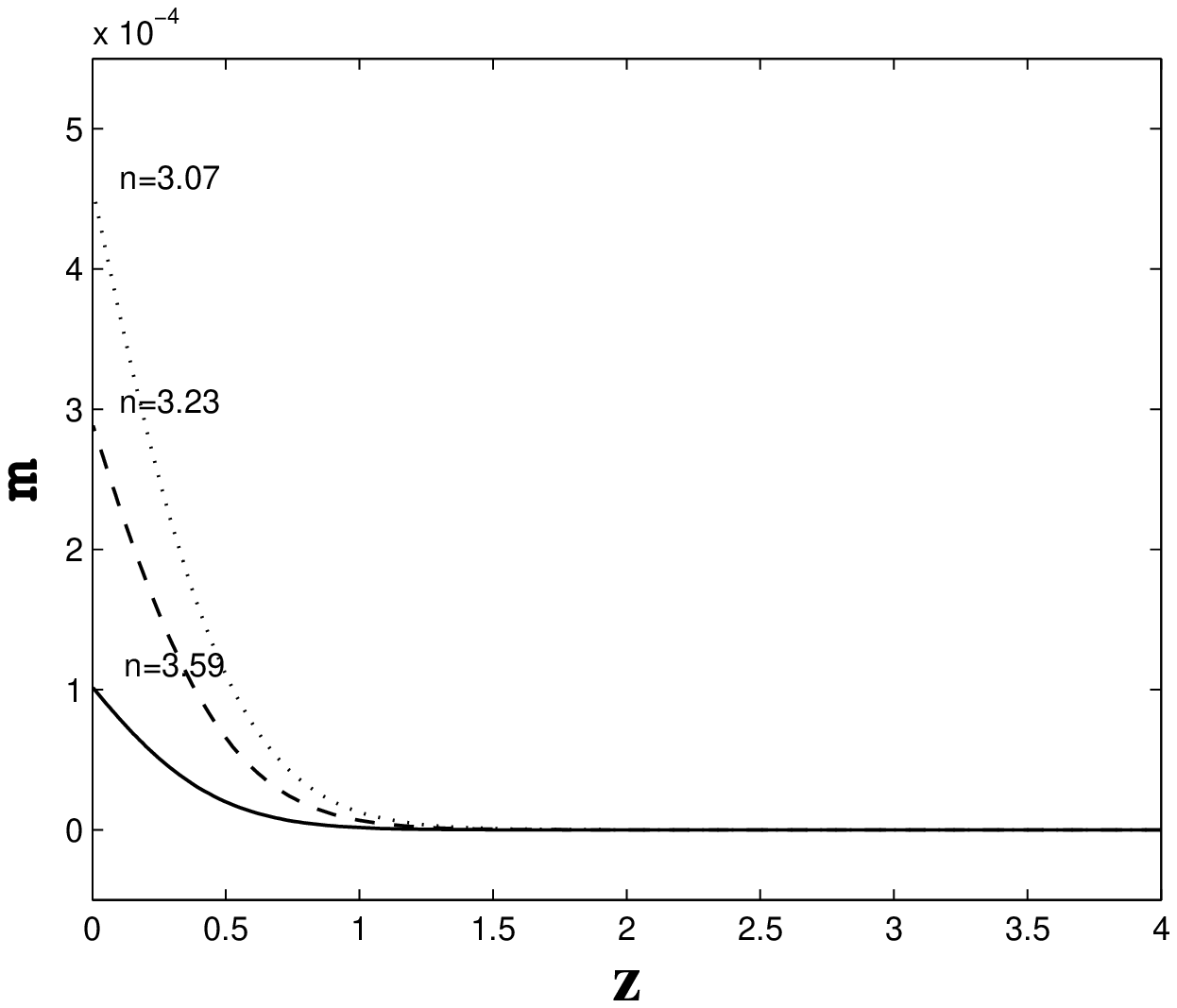}
\caption{\label{fig8}
The evolution of perturbations for the model
$f(R) = R-\lambda R_{c} [1-
\left(1+R^2/R_c^2\right)^{-n}]$
with $\lambda=2.5$.
The left hand panel depicts $s=\delta'_m/\delta_m$
versus the redshift $z$ for the mode $k/a_0H_0=600$
with three different values of $n$.
The right hand panel shows the evolution of the quantity
$m$ with respect to $z$ for $k/a_0H_0=600$. }
\end{figure}

In Fig.~\ref{fig8} we plot, for  mode $k=600a_0H_0$,
the evolution of $s$ and $m$ for $\lambda=2.5$
with three different values of $n$.
When $n=3.07$ the critical redshift is given by
$z_c \sim 1.05$ with $m \sim 1 \times 10^{-5}$.
The rapid increase of $s$ occurs in the regime $\xi>1$,
after which the growth rate reaches a maximum value
$s_{\rm max} \sim 2$.
The present value of $m$ is found to be $m=4.5 \times 10^{-4}$,
which is an order of magnitude larger than its corresponding value
at $\xi=1$. Using the criterion (\ref{scon}), we obtain the constraints
$n > 3.23$ and $m(z=0)<2.9 \times 10^{-4}$ for $\lambda=2.5$.
The present value of $m$ in this model is
one order of magnitude larger than the corresponding
values in the constant $m$ models as well as
$f(R)=R-\lambda R_c (R/R_c)^{\beta}$ model.

%---------------------------------------------------
%----------------------------------------------------

We also find that in contrast to the
model $f(R)=R-\lambda R_c (R/R_c)^{\beta}$ the constraints
on $n$ for the Starobinsky model are sensitive to the
values of the parameter $\lambda$.
For larger values of $\lambda$ the constraints on $n$ is weaker.
For example, for $\lambda=10$ and $\lambda=50$ we find the corresponding
constraints on $n$ imposed by (\ref{scon}) to be $n>1.74$
($m(z=0) \sim 1.4 \times 10^{-4}$) and
$n>1.09$ ($m(z=0) \sim 1.1 \times 10^{-4}$) respectively.
This can be understood in the following way.
When $\lambda$ is increased, we obtain a larger ratio $R_1/R_c$ from
Eq.~(\ref{lamratio}), which also leads to a larger ratio $R/R_c$
in the past. Then from Eq.~(\ref{mevo2}) a smaller value of $n$
is sufficient to realize the condition $|m| \ll 1$.
It can also be seen from the form of the
action that the values of $R_{c}$ can also affect the constraints on $n$.
We find that for small $\lambda$ values, $R_c$ has a small effect
on the constraint, whereas for large values of $\lambda$ the
affect of changing $R_c$ is negligible.

{}From Eq.~(\ref{mevo2}) we find that $m$ can be negative
for $-1/2<n<0$ (and $\lambda>0$) in the regime $R \gg R_c$.
When $n$ is close to $0$, the models are close to the model
$f(R)=R-\lambda R_c(R/R_c)^\beta$ discussed above.
We find that $s(z=0)$ is larger than $-1$ for
$|n|<9.3 \times 10^{-5}$, in which case we have
$|m(z=0)|<4.5 \times 10^{-5}$.
When $n$ is close to $-1/2$, Eq.~(\ref{mevo2})
seems to suggest that the models should be close to
the constant $m$ models.
However, care needs to be taken in this case since
$m$ changes sign from negative to positive
at $(R/R_c)^2=1/(2n+1)$ in the deep matter dominated
epoch.  As a result, for $n$ close to $-1/2$,
we numerically find that the growth rate
$s$ shows a rapid growth for $(R/R_c)^2<1/(2n+1)$.
Thus, in the limit $n \to -1/2$, the models
do not behave as constant $m$ models and they are
excluded observationally.

We have also analysed the model $f(R)=R-\lambda R_c \frac{(R/R_c)^{2n}}
{(R/R_c)^{2n}+1}$ ($n>0$) of Hu \& Sawicki \cite{Hu07}
and have found the constraints on the parameters $n$ and $m(z=0)$ to be
$n>3.33$ and $m(z=0)<2.15 \times 10^{-4}$ respectively for $\lambda=2.5$ and
$k/a_0H_0=600$, which are similar to the constraints derived above.

In summary, the present values of $m$ are constrained to be
$m(z=0) \lesssim 10^{-4}$ from the bound (\ref{scon})
in both Starobinsky and Hu \& Sawicki models.

%%%%%%%%%%%%%%%%%%%%%%%%%%%%%%%%%%%%%%%%%%%%%%%%%%%%%%%%%%%%%%%%%%%%%%%%%%%%%%%%
\section{Conclusions}
%%%%%%%%%%%%%%%%%%%%%%%%%%%%%%%%%%%%%%%%%%%%%%%%%%%%%%%%%%%%%%%%%%%%%%%%%%%%%%%%
\label{s6}

We have made a detailed study of the evolution of
density perturbations in $f(R)$ gravity theories in both metric
and Palatini formalisms and employed them to
study the viability of models in each case.
To study the viability of concrete models
we considered three sets of constraints, provided by
the background cosmological evolution, local gravity
experiments and the evolution of matter density perturbations
respectively.

We began by considering the
cosmological and local gravity constraints.
For models satisfying these constraints, we proceeded
to study the additional constraints provided by the
evolution of density perturbations
to further constrain the model parameters as well as their
deviation from the $\Lambda$CDM model.

The $f(R)$ theories in the metric formalism are equivalent to
generalised Brans-Dicke theories with a scalar-field potential $V(\phi)$
and Brans-Dicke parameter $\omega_{\rm BD}=0$.
The presence of the field potential, allows
the construction of $f(R)$ models that satisfy the
local gravity constraints under the use of a chameleon
mechanism. We find that for typical models of the forms
(\ref{model1}) and (\ref{model2})
to satisfy the cosmological and local gravity constraints,
the parameter $m$ is required to be much smaller than unity
during the radiation and matter eras
but can grow to values of order of $0.1$ in the accelerated epoch.
Models in the metric formalism also suffer from an additional fine tuning
due to the presence of scalaron oscillating modes
(which is absent in the Palatini case).
Finally, to be stable these theories require $f_{,RR}$ to be positive.

On the other hand, the $f(R)$ theories in the Palatini formalism
correspond to generalised Brans-Dicke theories with a
scalar-field potential $V(\phi)$ and Brans-Dicke parameter
$\omega_{\rm BD}=-3/2$.
This makes these theories special in the sense that
the oscillating scalar degree of freedom (scalaron) is
absent in these theories and therefore the corresponding
fine tuning to the metric case does not exist.
Also unlike the case of the metric formalism, there is in this case
no notion of field mass $M$ that determines an interaction length
mediated by a fifth force. Thus the LGC for these
theories need to be analyzed
separately in contrast to theories with
$\omega_{\rm BD} \neq -3/2$.
The main condition required in this case in order to
satisfy the LGC is that $|m|$
is smaller than the order of unity.
Moreover the requirement for the cosmological viability
in the Palatini formalism is not severe compared to the metric case.
Thus in contrast to the metric case,
to satisfy the cosmological and local gravity constraints,
we do not require vanishingly small values of $m$
during radiation and matter dominated epochs
and furthermore $f_{,RR}$ can be negative in this case.
As a result, even models of the type
$f(R)=R-\mu^{2(n+1)}/R^n$ with $n>0$ can be allowed
at the background level unlike the metric case.

We then studied the constraints provided by the
evolution of density perturbations in each case.
In the case of the metric formalism we derived the
equations for matter perturbations
under sub-horizon approximations in several different gauges.
In regimes $M^2 \gg k^2/a^2$ (i.e., $\xi \ll 1$), we
found the approximate perturbation equations
to be valid even without using sub-horizon approximations,
provided that the scalaron mode is not dominant relative to
the matter-induced mode.
This is a consequence of the fact that, when $M^2 \gg k^2/a^2$,
the evolution of perturbations mimics that in General Relativity.
After the Universe enters the regime $M^2 \ll k^2/a^2$,
the modes are inside the Hubble radius due to the fact
that the condition $M^2 \gg R$ is required for the compatibility with LGC.
Thus, for the models that satisfy LGC,
as long as the scalarons do not dominate over the
matter-induced mode, approximate perturbation
equations are valid even for the modes that initially lie
outside the Hubble radius.
In the Palatini case the approximate equations
are even more reliable because of the absence of scalarons.

In the metric formalism, most viable $f(R)$ models take
the form $m(r)=C(-r-1)^p$ ($p>1$) in the regimes where the Ricci
scalar is larger than the order of the present cosmological value.
In these models, the modes relevant to the observed matter
power spectrum correspond to the regimes $M^2 \gg k^2/a^2$ with
the growth rate $s=\delta_m'/\delta_m=1$
at the beginning of the matter era.
These models typically enter the regime $M^2 \ll k^2/a^2$ during the
matter era in which the growth rate of matter perturbations
is given by $s=1.186$. If we use the present observational bound
$s \lesssim 1.5$, we do not obtain strong constraints on these models.
However, since the transition time at $k/a=M$ depends upon the
mode $k$, there is a difference in the spectral indices
between the matter power spectrum and the CMB spectrum
[see Eq.~(\ref{deln})].
If we take the bound $\Delta n <0.05$,
the models with $p \ge 5$ are allowed.
The present value of the parameter $m$ is constrained
to be $m(z=0) \lesssim 10^{-1}$.
Thus, while $m$ needs to be negligibly small during the radiation and
matter eras, one can have appreciable deviation from the
$\Lambda$CDM model around the present epoch.

In the Palatini formalism the approximate matter perturbation equations
are valid even for super-Hubble modes,
for models satisfying LGC ($|m| \ll 1$).
If $m$ is positive, there is a strong amplification of $\delta_m$
in the regime $\xi \gg 1$, whereas if $m<0$ the matter perturbation
exhibits a damped oscillation for $|\xi| \gg 1$.
When the quantity $m$ evolves as $m \propto t^{2p}$
during the matter era,
we have analytically estimated the growth rate $s$
in both positive and negative $m$ cases
[see Eqs.~(\ref{delses}) and (\ref{delms2}), respectively].
{}From the requirement that the Universe does not enter
the regimes  $|\xi|>1$ during the matter era, we obtain
the constraint $|m(z)| \lesssim (a_0H_0/k)^2$ for $z>{\cal O}(1)$.
While this is a good criterion to avoid non-standard evolution of
matter perturbations, one needs to carry out a more detailed analysis
to place constraints on the quantity $m$ for each $f(R)$ model.
When $m$ is positive, we have obtained the constraint
$m \lesssim 10^{-5}$ by considering the modes $k$
relevant to the matter power spectrum.
We also studied the evolution of perturbations for the models
$f(R)=R-\lambda R_c (R/R_c)^{\beta}$ and
$f(R) = R-\lambda R_{c} [1-(1+R^2/R_c^2 )^{-n}]$.
For these models we found the constraints
$m(z=0) \lesssim 10^{-5}$ and $m(z=0) \lesssim 10^{-4}$,
respectively, from the requirement $s \lesssim1.5$.
Thus, unlike the metric case, the deviation from
the $\Lambda$CDM model at the present epoch is small
even when $m$ grows from the matter era to the accelerated epoch.
This situation does not change for negative values of $m$.

In summary, for viable models in the metric formalism, the quantity $m$
is constrained to be very much smaller than the order of
unity during the matter era from LGC,
but it can grow to the order of 0.1 around the present epoch.
In the Palatini formalism,  LGC and background cosmological constraints
do not place strong bounds on $m$ (only requiring
$|m| \lesssim 10^{-1}$), but the density
perturbations can provide stringent constraints:
$|m| \lesssim 10^{-5}$-$10^{-4}$.
Thus in the Palatini case the $f(R)$ theories are hardly
distinguishable from the $\Lambda$CDM model even
at the present epoch.
This follows from a peculiar evolution of the matter perturbations
in the Palatini case, in the regime $|\xi|>1$,
that exhibits rapid growth (when $m>0$)
or damped oscillations (when $m<0$).

While the constraints obtained here are sufficient to
give the orders of magnitude of the allowed model parameters,
it will be of interest to obtain more precise constraints
by using recent and upcoming observational
data including large scale structure,
CMB, Supernova Ia, gamma ray bursts and weak lensing.

%%%%%%%%%%%%%%%%%%%%%%%%%%%%%%%%%%%%%
\section*{ACKNOWLEDGEMENTS}

We thank Tomi Koivisto and Nikolay Koshelev for pointing out several typos.
ST thanks Masahiro Takada for fruitful discussions and
kind hospitality during his stay at Tohoku University.
We also thank Nikolay Koshelev for pointing out several typos
found in Ref.~\cite{Koivisto}.
ST is supported by JSPS (Grant No.\,30318802).
KU and RT would like to thank Karim Malik for usefull
discussions. KU is supported by the
Science and Technology Research Council (STFC).
RT would like to thank University of Rikkyo, Tokyo, and
Gunma National College of Technology
for kind hospitality during his stay where part of this work
was done.
%%%%%%%%%%%%%%%%%%%%%%%%%%%%%%%%%%%%%%%%%%%%%%%%%%%%%%%%%%%%%%%%%%%%%%%%%%%%%%%%%

\section*{Appendix A: The equation for matter perturbations in the comoving gauge in the
Palatini formalism}
%%%%%%%%%%%%%%%%%%%%%%%%%%%%%%%%%%%%%%%%%%%%%%%%%%%%%%%%%%%%%%%%%%%%%%%%%%%%%%%%

In this appendix we present the exact matter perturbation equation
in the comoving gauge in the Palatini formalism.
As in the metric case this equation needs to be solved
simultaneously with the background
equations (\ref{Pabe1})-(\ref{Pabe2}).
Unlike the metric case, however, it is not easy in this case
to find dimensionless variables in terms
of which both sets of equations close.
As a result we proceed to integrate the equations directly.
Using the background equations and ignoring the radiation,
the perturbation equation (\ref{delmpa}) can be written as
\begin{equation}
\label{CLdelmpa}
P_1\delta_m^{(v)''}+P_2\delta_m^{(v)'}+
P_3\delta_{m}^{(v)}=0\,,
\end{equation}
where the coefficients $P_{1}, P_2, P_{3}$ are given by
\begin{eqnarray}
\label{p1}
P_{1}&=&\left(1-\frac{3J}{2F}\right)H^2\,, \\
\label{p2}
P_{2}&=&\left(2+\frac{15J}{2F}\right)H^2+\left(1-\frac{3J}{2F}\right)\dot{H}+
\left(\frac{1}{2}+\frac{6J}{F}\right)\frac{\dot{F}H}{F}
-\frac{3J}{F}\frac{H\dot{F_{,R}}}{F_{,R}}-\frac{3J}{F} \frac{H\dot{m}}{1-m}\,, \\
\label{p3}
P_{3}&=&\frac{-\rho_{m}}{2F}-\frac{J}{2F}\left(6H^2+6\dot{H}+\frac{3\dot{F}^2}{F^2}-R+
\frac{k^2}{a^2}\right)-\frac{J}{2F}\left(3H-\frac{6\dot{F}}{F}\right)
\left(\frac{\dot{F_{,R}}}{F_{,R}}-3H-\frac{\dot{F}}{F}+\frac{\dot{m}}{1-m}\right) \nonumber  \\
&-&\frac{3J}{2F}\bigg[
\left(\frac{\dot{F_{,R}}}{F_{,R}}-3H-\frac{\dot{F}}{F}+\frac{\dot{m}}{1-m}\right)^2
+\frac{\ddot{F_{,R}}}{F_{,R}}
-\left(\frac{\dot{F_{,R}}}{F_{,R}}\right)^2-3\dot{H}-
\frac{\ddot{F}}{F}+\left(\frac{\dot{F}}{F}\right)^2
+\frac{\ddot{m}}{1-m}+\left(\frac{\dot{m}}{1-m}\right)^2 \bigg]\,,
\nonumber \\
\end{eqnarray}
where $J$ is defined as
\begin{equation}
\label{J}
J \equiv \frac{F_{,R}}{F}\frac{\rho_{m}}{1-m}.
\end{equation}
All the terms in the coefficients
$P_{1}, P_2, P_{3}$ can be expressed in terms of the scale factor $a$ (or
equivalently $N$), which thus allows Eq.~(\ref{CLdelmpa}) to close and
be readily integrated numerically.

%%%%%%%%%%%%%%%%%%%%%%%%%%%%%%%%%%%%%%%%%%%%%%%%%%%%%%%%%%%%%%%%%%%%%%%%%%%%%%%%%
\section*{Appendix B: The equation for matter perturbations
in the uniform density gauge in the Palatini formalism}
%%%%%%%%%%%%%%%%%%%%%%%%%%%%%%%%%%%%%%%%%%%%%%%%%%%%%%%%%%%%%%%%%%%%%%%%%%%%%%%%

In the uniform density gauge ($\delta \rho_m=0$)
we have $\delta F=0=\delta R$ from Eq.~(\ref{delFpa})
and $\dot{v}^{(\delta)}=\alpha$,
$\kappa=3H\dot{v}^{(\delta)}+\frac{k^2}{a^2}v^{(\delta)}$
from Eqs.~(\ref{ma1d}) and (\ref{ma2d}).
Substituting these relations for Eq.~(\ref{pa4}) we obtain
\begin{eqnarray}
3 \left(H+\frac{\dot{F}}{2F} \right) \ddot{v}^{(\delta)}+
\left( 6H^2+6\dot{H}+\frac{3\ddot{F}}{F}+
\frac{3H\dot{F}}{F}-\frac{3\dot{F}^2}{F^2}
\right) \dot{v}^{(\delta)}+\frac{\dot{F}}{2F}
\frac{k^2}{a^2}v^{(\delta)}=0\,,
\end{eqnarray}
where we have used the relation
\begin{equation}
R=6(2H^2+\dot{H})+\frac{3}{F}
\left( \ddot{F}+3H\dot{F}-\frac{\dot{F}^2}{2F} \right).
\end{equation}

Then the matter perturbation, $\delta_m^{(\delta)}=3Hv^{(\delta)}$,
satisfies the following equation of motion
\begin{eqnarray}
\label{delmpaUD}
\ddot{\delta}_m^{(\delta)}+c_1 \dot{\delta}_m^{(\delta)}
+c_2 \delta_m^{(\delta)}=0\,,
\end{eqnarray}
where
\begin{eqnarray}
\label{c1UD}
c_1 &=& \frac{2H}{1+\dot{F}/2HF}
\left[ 1+\left(1-\frac{\dot{H}}{H^2}\right) \frac{\dot{F}}{2HF}
-\frac{\dot{F}^2}{2H^2 F^2}+\frac{\ddot{F}}{2H^2 F} \right]\,, \\
\label{c2UD}
c_2 &=& \frac{H^2}{1+\dot{F}/2HF}
\left[ -\frac{\ddot{H}}{H^3}-\frac{2\dot{H}}{H^2}+\frac{\dot{H}}{H^2}
\left( \frac{\dot{F}}{HF} \right)^2+
\frac{\dot{F}}{HF} \left( \frac{\dot{H}^2}{H^4}-\frac{\ddot{H}}{2H^3}
-\frac{\dot{H}}{H^2}+\frac{k^2}{6a^2H^2} \right)
-\frac{\dot{H}}{H^2} \frac{\ddot{F}}{H^2F} \right]\,. \nonumber \\
\end{eqnarray}
This agrees with the result derived in Refs.~\cite{Koivisto,Uddin}.

Let us obtain the approximate equation for matter perturbations
under the approximation (\ref{mcon}).
Taking note that the term $|\dot{F}/HF|$ is of the order of
$|m|$, the coefficients $c_1$ and $c_2$ are given by
\begin{eqnarray}
c_1=2H\,, \quad
c_2=H^2 \left[-\frac{\ddot{H}}{H^3}-\frac{2\dot{H}}{H^2}
+\frac{\dot{F}}{6HF} \frac{k^2}{(aH)^2}
\right]\,.
\end{eqnarray}

When we estimate the first two terms in the square bracket
of $c_2$, we use the following approximate relations
\begin{eqnarray}
\label{apprho}
2F\dot{H} \simeq -\rho_m\,, \quad
2F\ddot{H} \simeq 3H\rho_m\,,
\end{eqnarray}
which come from Eqs.~(\ref{Pabe1}) and (\ref{Pabe2}).
From Eq.~(\ref{dotR}) we have
\begin{eqnarray}
\dot{F}=-\frac{3\rho_m F_{,R}H}{F-RF_{,R}}\,.
\end{eqnarray}
Using these relations, we find that the matter perturbation
satisfies the following approximate equation of motion
\begin{eqnarray}
\label{delmpa1}
\ddot{\delta}_m^{(\delta)}+2H\dot{\delta}_m^{(\delta)}
-\frac{\rho_m}{2F} \left(1+\frac{\xi}{1-m} \right)
\delta_m^{(\delta)} \simeq 0\,.
\end{eqnarray}
This is the same equation as the one in the longitudinal gauge (\ref{delmpa3}).

%%%%%%%%%%%%%%%%%%%%%%%%%%%%%%%%%%%%%%%%%%%%%%%%%%%%%%%%%%%%%%%%%%%%%%%%%%%%%%%%%
\section*{Appendix C: Equations convenient for numerical simulations in
The Palatini formalism}
%%%%%%%%%%%%%%%%%%%%%%%%%%%%%%%%%%%%%%%%%%%%%%%%%%%%%%%%%%%%%%%%%%%%%%%%%%%%%%%%

In this appendix we shall present the equations convenient for numerical
simulations. {}From Eqs.~(\ref{Pabe1}), (\ref{Pabe2}) and (\ref{dotR})
we obtain
\begin{eqnarray}
H^2=\frac{2\rho_m+FR-f}{6F\zeta}\,,~~~
{\rm where}~~~
\zeta=\left[ 1-\frac32 \frac{F_{,R}(FR-2f)}
{F(F_{,R}R-F)} \right]^2\,.
\end{eqnarray}
Introducing a dimensionless quantity
\begin{eqnarray}
\label{ydef}
y=\frac{FR-f}{6F\zeta H^2}\,,
\end{eqnarray}
we get the differential equation for $y$ \cite{FTS}:
\begin{eqnarray}
\label{ydif}
y'=y(1-y) \left[3+C(R)\right]\,,
\end{eqnarray}
where $C(R)$ is defined in Eq.~(\ref{CR}).

The following relations also hold
\begin{eqnarray}
\label{Palare1}
& & \frac{FR-f}{FR-2f}=-\frac{2y}{1-y}\,,\\
\label{Palare2}
& & \Omega_m \equiv
\frac{\rho_m}{3F\zeta H^2}=1-y\,.
\end{eqnarray}
Specifying the value of $y$, the initial Ricci scalar $R$
is determined by Eq.~(\ref{Palare1}).
Solving Eq.~(\ref{ydif}), we obtain $y$, $R$, $H$
and $\Omega_m$ from Eqs.~(\ref{Palare1}),
(\ref{ydef}) and (\ref{Palare2}).
The effective equation of state of dark energy is given by
\begin{eqnarray}
w_{\rm eff}=-y+\frac{\dot{F}}{3HF}
+\frac{\dot{\zeta}}{3H\zeta}-\frac{\dot{F}R}
{18F\zeta H^3}\,.
\end{eqnarray}
As long as the deviation from the $\Lambda$CDM model
is small ($|m| \ll 1$), we have $w_{\rm eff} \simeq -y_1$.

The perturbation equations (\ref{delmpa3}) and (\ref{Poipala}) are given by
\begin{eqnarray}
& &\delta_m''+\frac12 (1-3w_{\rm eff})\delta_m'
-\frac32 \zeta (1-y) \left( 1+\frac{\xi}{1-m}
\right)\delta_m \simeq 0\,,\\
& & \Phi_{\rm eff} \simeq -\frac32 \left( \frac{aH}{k}
\right)^2 \zeta (1-y) \delta_m\,.
\end{eqnarray}
Although we solve exact perturbation equations,
the above perturbation equations are
found to be very accurate.

%----------------------------------------------------%
\bibliographystyle{unsrt}

\begin{thebibliography}{10}
\bibitem{comment}
In Ref.~\cite{BLi} it was shown that the combined
analysis using the SDSS, CMB and Supernovae Ia data gives more
stringent constraints: $\beta \lesssim 10^{-6}$.
\end{thebibliography}

\end{document}